\documentclass{aa}

\usepackage{graphicx}
\usepackage{multirow}
\usepackage{xcolor}
\usepackage{color}
\usepackage{amsmath}    
\usepackage{txfonts}
\usepackage{hyperref}
\usepackage{multirow}
\usepackage{notes2bib}
\usepackage{stfloats}
\usepackage{float}
\restylefloat{table}

\hypersetup{colorlinks=true,linkcolor=[rgb]{1.,0.2,0.2},citecolor=[rgb]{0.1,0.4,1.},filecolor=[rgb]{0.7,0.2,0.2},urlcolor=[rgb]{0.0,0.2,1.}}
\definecolor{darkgreen}{rgb}{0.09, 0.45, 0.27}
\definecolor{amber(sae/ece)}{rgb}{1.0, 0.49, 0.0}
\defcitealias{Tortora+18_UCMGs}{T18}
\defcitealias{Scognamiglio20}{S20}
\defcitealias{Ferre-Mateu+17}{F17}
\defcitealias{Spiniello20_Pilot}{\INSPIRE\, Pilot}

\def\INSPIRE{\mbox{{\tt INSPIRE}}}

\newcommand{\Reff}{$\mathrm{R}_{\mathrm{e}\,}$}
\newcommand{\Mstar}{M$_{\star}\,$}

\newcommand{\kms}{km s$^{-1}$}
\newcommand{\Msun}{M$_{\odot}\,$}

\usepackage[normalem]{ulem}
\newcommand{\ppxf}{\textsc{pPXF}}

\begin{document} 

   \title{INSPIRE: INvestigating Stellar Population In RElics}
   \subtitle{II. First data release (DR1)}

   \author{C.~Spiniello\inst{\ref{oxf},\ref{inaf_naples}}\fnmsep\thanks{Corresponding author: {\tt chiara.spiniello@physics.ox.ac.uk}.} 
   \and C.~Tortora\inst{\ref{inaf_naples}}
   \and G.~D'Ago\inst{\ref{puc}}
   \and L.~Coccato\inst{\ref{eso}}
   \and F.~La Barbera\inst{\ref{inaf_naples}}
   \and A.~Ferr\'e-Mateu\inst{\ref{iccub}}
   \and C.~Pulsoni\inst{\ref{mpe}}
   \and M.~Arnaboldi\inst{\ref{eso}}
   \and A.~Gallazzi\inst{\ref{inaf_arcetri}}
   \and L.~Hunt\inst{\ref{inaf_arcetri}}
   \and N.~R.~Napolitano\inst{\ref{inaf_naples},\ref{SunYat}}
   \and   M.~Radovich\inst{\ref{inaf_padova}}
    \and  D.~Scognamiglio\inst{\ref{bonn}} 
    \and M.~Spavone\inst{\ref{inaf_naples}}
   \and S.~Zibetti\inst{\ref{inaf_arcetri}}
   }

   \institute{Sub-Dep. of Astrophysics, Dep. of Physics, University of Oxford, Denys Wilkinson Building, Keble Road, Oxford OX1 3RH, UK\label{oxf} 
    \and
    INAF -  Osservatorio Astronomico di Capodimonte, Via Moiariello  16, 80131, Naples, Italy\label{inaf_naples}
    \and
    Institute of Astrophysics, Pontificia Universidad Cat\'olica de Chile, Av. Vicu\~na Mackenna 4860, 7820436 Macul, Santiago, Chile\label{puc}
        \and 
    European Southern Observatory,  Karl-Schwarzschild-Stra\ss{}e 2, 85748, Garching, Germany\label{eso}
    \and
    Institut de Ciencies del Cosmos (ICCUB), Universitat de Barcelona (IEEC-UB), E-02028 Barcelona, Spain\label{iccub} 
    \and 
    Max-Planck-Institut f\"{u}r  extraterrestrische Physik, Giessenbachstrasse, 85748 Garching, Germany\label{mpe}
   \and
    INAF - Osservatorio Astronomico di Arcetri, Largo Enrico Fermi 5, 50125, Firenze, Italy\label{inaf_arcetri}    
    \and
    School for Physics and Astronomy, Sun Yat-sen University, Guangzhou 519082, Zhuhai Campus, China\label{SunYat} 
    \and 
    INAF - Osservatorio astronomico di Padova, Vicolo Osservatorio 5, I-35122 Padova, Italy\label{inaf_padova}
        \and
    Argelander-Institut f\"{u}r Astronomie, Auf dem H\"{u}gel 71, D-53121 - Bonn, Germany\label{bonn}
}

   \date{Accepted for publication 21-07-2021 }

 
  \abstract
   {The INvestigating Stellar Population In RElics (\INSPIRE) is an ongoing project targeting 52 ultra-compact massive galaxies at $0.1<z<0.5$ with the X-Shooter@VLT spectrograph (XSH). These objects 
   are the ideal candidates to be `relics', massive red nuggets that have formed at high redshift ($z>2$) through a short and intense star formation burst, and then have evolved 
   passively and undisturbed until the present day. Relics provide a unique opportunity to study  
   the mechanisms of star formation at high-z.}
   {\INSPIRE\, is designed to spectroscopically confirm and fully characterise a 
   large sample of relics, computing their number density 
   in the redshift window $0.1<z<0.5$ for the first time, thus providing a benchmark for cosmological galaxy formation simulations.  In this paper, we present the \INSPIRE\, Data Release (DR1), comprising 19 systems with observations completed in 2020. }
   {We use the methods already presented in the \INSPIRE\, Pilot, but revisiting the 1D spectral extraction. For the 19 systems studied here, we obtain an estimate of the stellar velocity dispersion, fitting the two XSH arms (UVB and VIS)  separately at their original spectral resolution to two spectra extracted in different ways. We estimate [Mg/Fe] abundances via line-index strength and mass-weighted integrated stellar ages and metallicities with full spectral fitting on the combined (UVB+VIS) spectrum.}
   {For each system, different estimates of the velocity dispersion always agree within the errors.  
   Spectroscopic ages are very old for 13/19 galaxies, in agreement with the photometric ones, and metallicities are almost always (18/19) super-solar, confirming the  mass--metallicity relation. The [Mg/Fe] ratio is also larger than solar for the great majority of the galaxies, as expected. We find that ten objects formed more than $75\%$ of their stellar mass (M$_{\star}$) within 3 Gyr from the big bang (BB) and classify them as relics. Among these, we identify four galaxies that had already fully assembled their M$_{\star}$ by that time and are therefore `extreme relics' of the ancient Universe.  Interestingly,  relics,  overall, have a larger [Mg/Fe] and a more metal-rich stellar population. They also have larger integrated velocity dispersion values compared to non-relics (both ultra-compact and normal-size) of similar stellar mass.} 
   {The INSPIRE DR1 catalogue of ten known relics 
   is  the largest  publicly available collection, augmenting the total number of confirmed relics  by a factor
of 3.3, and also enlarging the redshift window. The resulting lower limit for the number density of relics at $0.17<z<0.39$ is $\rho \sim 9.1 \times 10^{-8} \, \text{Mpc}^{-3}$. }  


   \maketitle
%


\section{Introduction}
\label{sec:intro}
Relic galaxies are defined as massive ($\mathrm{M}_{\star}\ge 6 \times 10^{10}$ \Msun) 
very compact (\Reff $\,\le 2$ kpc) objects that formed early on in cosmic time ($z>2$), through a short and intense star formation episode ($\tau \sim 100 {\rm Myr , \,\, SFR} \ge 10^{3} {\rm M}_{\odot} {\rm yr}^{-1}$), and then evolved passively and undisturbed until the present day. 
They experienced very little or completely missed the second phase of the so-called two-phase formation scenario which is often used to explain the mass assembly of the local, massive early-type galaxies (ETGs;  \citealt{Naab+09,Oser+10,Hilz+13,Rodriguez-Gomez+16}). As such, the great majority of their stellar mass is composed by an `in-situ', old stellar population with super-solar [Mg/Fe] abundance. 
The large [Mg/Fe] value is generally attributed to the quenching of star formation before Type Ia supernova explosions are able to pollute the interstellar  medium with iron \citep[e.g.][]{Thomas+05, Gallazzi+06, Gallazzi+14}. This stellar component generally ends up in the innermost core of 
massive ETGs at low redshift and in the local Universe \citep{Navarro-Gonzalez13, Pulsoni20}. A recent analysis based on the Illustris TNG100 simulation presented in \citet{Pulsoni20} demonstrated that the fraction of ETGs with a compact progenitor increases with increasing stellar mass and that at M$_{\star}> 10^{11}$M$_\sun$,  this fraction is as high as 80\%. However, only $\sim20$\% of present-day massive ETGs still have this pristine in situ component preserved in their innermost region. This mainly depends on the merger history of the system.  
The more massive the galaxy, the more the accreted component starts to dominate the mass and light budgets at smaller radii \citep[e.g.][]{Seigar07,Cooper13, Rodriguez-Gomez+16,Spavone17, Spavone20, Pulsoni20, Remus21}. This
significantly complicates any study of the high-z baryonic processes that formed the pristine\ `in-situ' component of giant ETGs \citep[e.g.][]{Naab+14,Ferre-Mateu+19}.  Moreover, the number density of relic galaxies as a function of redshift predicted by simulations critically depends on the different assumptions and ingredients used to calculate it (e.g. \citealt{Naab+09,Oser+10, Oser+12, Oogi13,Quilis_Trujillo13,Zolotov15,Wellons16,Pulsoni20}).
For all these reasons, relics are a niche in the star formation--mass/size parameter distribution of galaxies 
that allows us to put precise constraints on the physical mechanisms that shaped the Universe and the galaxy mass assembly 
at early cosmic time. 

These systems are unfortunately very rare. Only by scanning a wide portion of the night sky with multi-band photometric surveys, followed up by a large spectroscopic confirmation program, will it be possible to build a statistically valid sample of relics at $z < 0.5$, and be able to constrain their number density and redshift evolution. 
This is indeed the approach we undertook in the last few years within the Kilo Degree Survey (KiDS, \citealt{Kuijken11}) which, starting with DR4 \citep{Kuijken19_KIDSDR4}, is also complemented by near-infrared (NIR) data from the VISTA Kilo Degree Infrared Galaxy Survey (VIKING, \citealt{Sutherland12, Edge13}). Thanks to the high image quality (spatial resolution of $0.21\arcsec$/pixel, and a median r-band seeing of $\sim0.65\arcsec$) and wide coverage ($1350$ deg$^2$ in $ugrizYJHKs$) of KiDS+VIKING, we have built the largest sample ever constructed of photometrically selected ultra-compact massive galaxies (UCMGs, $\sim1000$ objects, \citealt[][hereafter T18]{Tortora+18_UCMGs}). 
We subsequently obtained low-resolution and low signal-to-noise ratio (S/N) spectra from a multi-site, multi-telescope spectroscopic program in order to infer the redshift and confirm the extra-galactic massive and compact nature of $\sim100$ UCMGs (\citetalias{Tortora+18_UCMGs} and \citealt[][hereafter S20]{Scognamiglio20}). 
The INvestigating Stellar Population In RElics   is the natural next step. 
Among these confirmed UCMGs, we now need to identify the galaxies whose stellar mass is almost entirely been assembled through a single, very short star formation episode that occurred only a few gigayears after the big bang (BB). 
To this end, high-resolution and high S/N spectra are essential, 
 as we demonstrated from the pilot program published as \citet[][hereafter \INSPIRE\, Pilot]{Spiniello20_Pilot}

In this paper, we introduce the European Southern Observatory (ESO) X-Shooter (XSH, \citealt{Vernet11}) spectroscopic Large Program (LP) \INSPIRE\, and release its data up to March 2020.  The data comprises the first 19 objects with complete observations. We also present the results based on the kinematical and stellar population analysis of 
UVB and VIS spectra extracted from two different methods. The first one uses a fixed aperture encapsulating  50\% of the light of the object, and the second sums up all the light from the slit, but gives more weight to the pixels with larger fluxes. For each of the 19 UCMGs, we report the velocity dispersion values computed from the UVB and VIS arms independently, finding a fair agreement within the two. We obtain a rough estimate of the [Mg/Fe] abundance from line-index measurements and finally constrain the mass-weighted age and metallicity of the stellar populations from full spectral fitting. This allows us to constrain the mass assembly of the galaxies as a function of time, the key parameter for determining whether or not they are relics. 

The paper is organised as follows. The main scientific aims and goals of \INSPIRE\, are described in Section~\ref{sec:survey_presentation}, the observations and data strategy of the ESO LP are described in Section~\ref{sec:LP_presentation}, and the data reduction and 1D extraction of the systems released as part of this \INSPIRE\, DR1 are described in Section~\ref{sec:data_analysis}. The kinematical analysis and results are presented in Section~\ref{sec:kinematics} and the stellar population analysis and results are given in Section~\ref{sec:stelpop}. In Section~\ref{sec:results} we present our results, including the `relic confirmation'.  In Section~\ref{sec:conclusions}, we present our conclusions, and finally, in Section~\ref{sec:future}, we give a brief overview on the future \INSPIRE\  data releases and scientific objectives. 

Throughout the paper, we assume a standard $\Lambda$CDM cosmology with $H_0=69.6$ \kms Mpc$^{-1}$, $\Omega_{\mathrm{vac}} = 0.714$ and $\Omega_{\mathrm{M}} = 0.286$ \citep{Bennett14}.

\section{Scientific aims and objectives of \INSPIRE}
\label{sec:survey_presentation}

The primary goal of the \INSPIRE\, Project is 
to build the first large catalogue of relics at $0.1<z<0.5$, in order to put constraints on the first phase of the mass assembly of ETGs in the Universe and on the mechanisms responsible for their size evolution in cosmic time. 
In the following, we provide an overview of the main scientific goals of the project.

\textbf{1. Studying the mass assembly and galaxy growth by measuring the number density of relics.}\\
From a theoretical point of view, different high-resolution cosmological simulations predict a different relic number density evolution with redshift.  
Among the state-of-the-art semi-analytical models and simulations \citep{Quilis_Trujillo13, Wellons16, Furlong17}, the fraction of relics that survive until today varies from 0\% to 15\% and heavily depends on the different assumptions and ingredients used to calculate it and even more critically on the physical processes shaping the size growth and mass evolution of galaxies, such as the relative importance of major and minor galaxy merging, stellar winds, and AGN feedback (e.g. \citealt{Naab+09, Oser+10, Oser+12, Oogi13, Wellons16}). 
Therefore, a precise census of  
the number of relics in distinct redshift bins  
is a crucial and a very valuable way to test different galaxy evolution models and thus to disentangle between possible physical scenarios driving the formation and size-evolution of galaxies. With the \INSPIRE\, catalogue, we will  bridge the gap between the only three confirmed local relics ($z<0.03$, \citealt{Trujillo+14},  \citealt[hereafter F17]{Ferre-Mateu+17}) and the high-z red nuggets, putting statistical constraints (at least a lower limit) on how many relics exist in the Universe at each redshift. A full comparison between the \INSPIRE\, galaxies  with simulated UCMGs in IllustrisTNG will be addressed in a separate paper (Pulsoni et al., in prep.). This will allow us to shed light on the formation scenario and mass assembly of these extremely compact, passive, and massive objects, and to understand the conditions that allowed them to survive intact until the present day without undergoing any accretion event.

\textbf{2. UCMGs and relics in different environments.}\\
Thanks to the large and uniform coverage of the KiDS Survey from which the \INSPIRE\, targets are taken, we have the novel possibility to study the environment in which relics preferentially live.  
On one hand, some works have claimed that UCMGs could be more common in denser environments \citep{Stringer+15_compacts, PeraltadeArriba+16} where gas stripping prevents secondary star formation events.  The high velocity dispersion in such over-densities would avoid further merging, as long as the galaxy entered the cluster early enough \citep[e.g.][]{Poggianti+13_low_z}.  However, very recently it has been shown that the relative fraction of ultra-compact and normal-sized passive and massive galaxies depends only mildly on the global environment \citep{Tortora20}. It has been proven that,  
while denser environments slightly penalise the survival of relics, the number of massive passive galaxies with any size is much larger in denser environments and thus it is much easier to find UCMGs there \citep{Damjanov+15_env_compacts, Baldry21}. Focussing in particular on relics, instead, \citetalias{Ferre-Mateu+17} highlighted that a kind of `degree of relicness' might exist, in the sense that some of the morphological and stellar characteristics of the relics might be more or less extreme in different environments. A hint in this direction seems to also be present  in the three galaxies analysed in the \citetalias{Spiniello20_Pilot} but this can only be fully confirmed with a larger sample of relics.  
When \INSPIRE\, is completed, we will obtain, for each object, the  probability of belonging to a cluster of galaxies using the Adaptive Matched Identifier of Clustered Objects (AMICO,  \citealt{Maturi19}). This  algorithm identified $\sim8000$ candidate galaxy clusters in the redshift range 0.1 < z < 0.8 with a purity of $\sim95$\%. \\
\indent \textbf{3. The low-mass end of the initial mass function slope}.\\
Another objective of the \INSPIRE\, Project, which is made possible by the very large wavelength range of XSH is a direct spectroscopical constraint on the slope of the stellar initial mass function (IMF). 
All three local relics of \citetalias{Ferre-Mateu+17} need a bottom-heavy IMF  up to a few effective radii, that is, they have a fraction of stars with M $<0.5M_{\odot}$ which is at least a factor of two larger than the one measured in the Milky Way (\citealt{Martin-Navarro+15_IMF_relic}). 
The same result has also been found in very massive local ETGs of normal size.  However, in this latter case, the bottom-heavy IMFs are only measured in the innermost region \citep{Martin-Navarro+15_IMF_variation, Sarzi+18, Parikh+18,LaBarbera+19, Barbosa21_letter}, where the relic component is supposed to dominate the light budget \citep{Navarro-Gonzalez13,Barbosa21_letter}. Thus, a possible scenario to explain these observations is that the IMF might have been bottom-heavy during the early stages of galaxy formation when the Universe was much more dense and richer in hot gas and only if stars formed through a very intense (${\rm SFR} \ge 10^{3} {\rm M}_{\odot} {\rm yr}^{-1}$) and very short ($\tau\sim 100 {\rm Myr}$) burst \citep{Smith2020ARA&A}. 
In relics, where  almost the totality of the stellar mass was formed during the first phase, we should measure a very steep IMF slope even  from an integrated spectrum covering a large portion of the galaxy size.  
Inferring the IMF slope from spectra is, however, a very challenging task, 
because the degeneracies between the stellar population parameters are large and the contribution of low-mass stars to the overall galaxy light is very small (see e.g. Fig.~2 in \citealt{Conroy_vanDokkum12a}). 
The XSH spectrograph is the perfect instrument to simultaneously constrain (from the same spectrum) stellar age, metallicity, elemental abundances ([Mg/Fe]), and the IMF slope thanks to its wide wavelength range. 
Specifically, it covers blue lines, which are needed to estimate age, metallicity, and elemental abundances (e.g. CaK, CaH, H$\beta$, Mg$_{b}$, Fe5270, Fe5335, NaD), and redder gravity-sensitive features (TiO and CaH molecular bands, as well as Na, Ca, and Fe lines in the NIR) necesasary to break the stellar population degeneracies and obtain a solid estimation of the low-mass end of the IMF\footnote{M-dwarfs stars contribute more to the light in the NIR, 
which is why it is 
easier to constrain the low-mass end of the IMF in this wavelength range.}. 
High-S/N spectra covering the full optical to IR range can simply not be obtained with any other instrument  available on any telescope, making the XSH@VLT instrument unique for these detailed studies. Thus, with XSH, we will be able to put firm constraints on the IMF slope, while at the same time inferring light-weighted integrated age, metallicity, and elemental abundance. 

\begin{figure}
\includegraphics[width=\columnwidth]{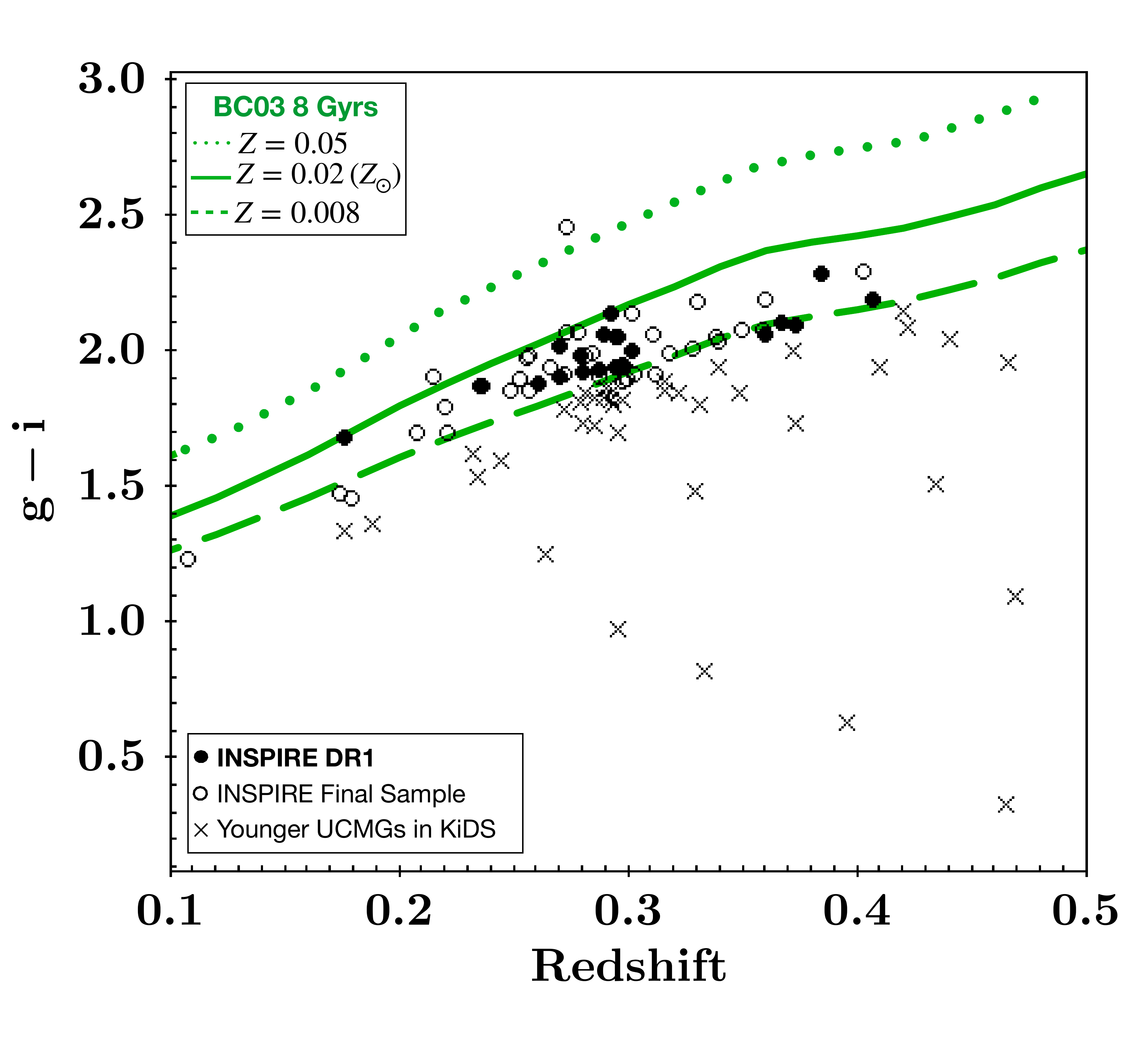}
\caption{The $g-i$ colour as a function of redshift for the \INSPIRE\, final sample (empty circles) and for the objects released as part of this DR1 (filled circles).  Lines are single-burst stellar population models \citep{Bruzual+03} with solar (Z$=0.02$, solid lines),  subsolar (Z$=0.008$, dashed), and super-solar (Z$=0.05$, dotted) metallicity and ages of 8 Gyr. Crosses are younger objects from the confirmed 
KiDS UCMGs spectroscopic catalogue (\Reff$<2$ kpc and  \Mstar$>6\times10^{10}$\Msun).} 
\label{fig:color_selection}
\end{figure}

In summary, the \INSPIRE\, Project is a fundamental step forwards in shedding light on  the mechanisms responsible for the dramatic growth in size measured from $z\sim1$ to the present-day Universe, and thus in better understanding the formation and evolution of the most massive galaxies in our local Universe. 
The first step to reach all of the above-mentioned scientific goals is to study  the stellar populations of UCMGs in great detail, and thus find the ones that are indeed relics. 
This is the main purpose of this paper, presenting the first Data Release of the spectra observed as part of the ESO LP that we describe in the following section. 

\section{The ESO Large Program \INSPIRE}
\label{sec:LP_presentation}
The \INSPIRE\, dataset is obtained as part of an ESO LP (ID: 1104.B-0370, PI: C. Spiniello) 
to spectroscopically follow
up with XSH 52 UCMGs with redshift $0.1<z<0.5$, which are part of a dedicated KiDS project \citepalias{Tortora+18_UCMGs,Scognamiglio20}. 

According to the definition given in \citet{Trujillo+09_superdense} and used in  \citetalias{Tortora+18_UCMGs}, a galaxy is defined as UCMG if it has an effective radius \Reff$<1.5$ kpc and a stellar mass \Mstar$>8\times10^{10}$\Msun.  However, as different studies have adopted different thresholds for size and masses, and as we aim at finding a large number of relics, we slightly relax these criteria and consider all objects with \Reff$<2$ kpc and  \Mstar$>6\times10^{10}$\Msun as confirmed UCMGs. 

Starting from a catalogue of $117$ objects with structural parameters computed from $gri$ KiDS images \citep{Roy+18}, spectroscopic redshifts, and stellar masses retrieved from \citetalias{Tortora+18_UCMGs} and \citetalias{Scognamiglio20}, we select only those with  the $g-i$ broad band colour compatible with that of a stellar population with integrated age $\ge8$ Gyr (considering a solar, super-solar, and a subsolar metallicity).  
To this purpose, we use single-burst stellar population models from \citet{Bruzual+03} and the MAG\_AUTO magnitudes from the KiDS DR3 after correcting them for extinction. In addition, we discard objects for which spectra with good S/N are already available from the literature\footnote{We will analyse these spectra with the same tools presented here, hence constraining the relic nature of a few more UCMGs.}. Applying these selection cuts, we finally selected 52 galaxies for the XSH follow up. 
Figure~\ref{fig:color_selection} shows the $g-i$ colour as a function of redshift, in the windows where the candidates lay. 
Empty circles are the 52 galaxies we target with the \INSPIRE\, LP (final sample), while filled ones are the objects with complete observations released with this DR1. Finally, crosses are UCMGs that we discard because their colours are consistent with younger ages\footnote{This figure is not identical to the one shown in the \citetalias{Spiniello20_Pilot}  because, thanks to new information obtained on the candidates (from the $9$-band optical+NIR photometric catalogue released as part of KiDS DR4), and also due to the delay in the observations caused by the shutting down of the ESO telescopes during the COVID-19 pandemic, we have slightly amended the list of targets.}. 

\begin{figure}
\includegraphics[width=\columnwidth]{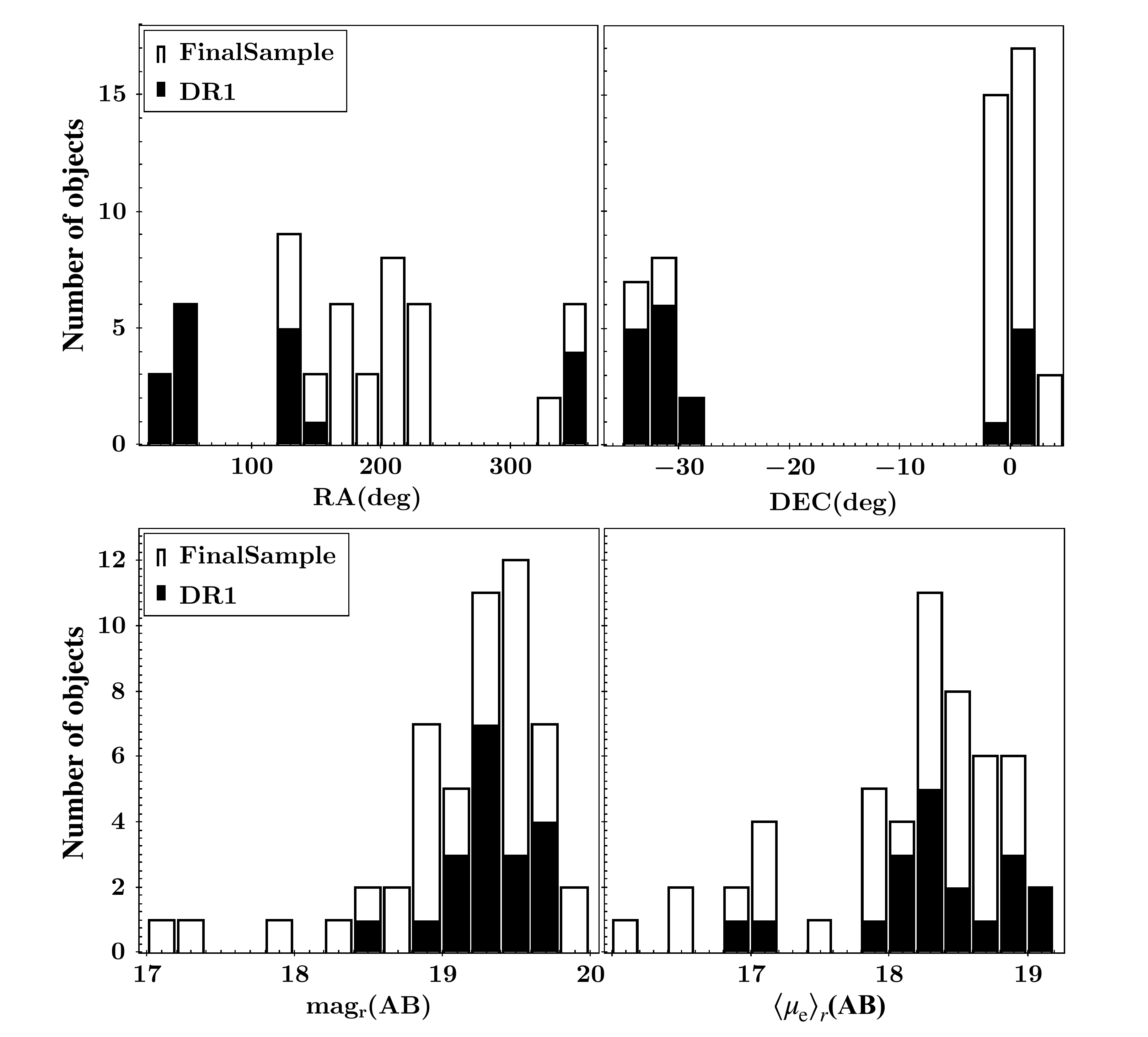}
\caption{Distribution of the \INSPIRE\, targets in right ascension and declination (\textit{upper panels}) and in $r$-band aperture magnitude (MAG\_AUTO\_r) and surface brightness $r$-band luminosity averaged within the effective radius (\textit{bottom panels}). The empty histograms show the distributions of the final \INSPIRE\, sample, while the filled ones represent the distributions of the objects in this first Data Release.}
\label{fig:obj_distrib}
\end{figure}

\subsection{Observations strategy and current status}
\label{subsec:data}
The observation strategy has been optimised to capitalise on relatively substandard observing conditions (seeing up to 1.2 arcsec, clear nights, grey lunar phase with fraction of lunar illumination $\le0.5$), allowing for easy scheduling of the objects into the observation queue. 
Moreover, the selected targets span a very wide range in right ascension (RA) and declination (DEC), as can be seen from the histograms in the upper panels of Figure~\ref{fig:obj_distrib}, with an optimum observing time spread over the full year. This makes service mode observations under an LP highly efficient. We also note that we have many systems with DEC$<-30$, which are perfect as `fillers' in nights with strong wind coming from the north.  
As for the previous figure, filled histograms show the distribution of galaxies with completed datasets (19 systems) that form this DR1, while empty ones refer to the whole \INSPIRE\, sample (52 systems).  

For all the observations, we set the slit widths to 1.6 arcsec in the UVB and 1.5 arcsec in VIS and NIR to ensure minimal slit loss.  Finally, we use a dithering scheme (NODDING MODE) with multiple frames where the galaxy is offset by a small amount from the centre of the slit to facilitate a proper sky subtraction. 

The final integration time on target has been driven by the high S/N ratio we need to reach in order to precisely constrain the stellar population parameters. In particular, because these systems are generally red, the exposure times are set to reach a S/N in the blue part of the optical that is high enough to infer stellar ages and break the degeneracies with other stellar population parameters. 
Based on our previous experience and on literature results \citep{CidFernandes14, Ge+18, Costantin19}, an integrated S/N $\ge15$ per \AA \, allows the recovery of the stellar population parameters through full spectral fitting with uncertainties smaller than $0.1-0.2$ dex, and thus the confirmation of the relic nature. We note that a higher S/N ($\ge50$ per \AA, based on the authors' experience) is instead required to properly infer the low-mass end slope of the IMF (see e.g. \citealt{LaBarbera+19}). We therefore plan to  stack the spectra of all the confirmed relics together in order to build a 1D spectrum with sufficient S/N for our purposes. 

We use the surface brightness luminosities ($\langle\mu_{\mathrm {e}}\rangle$) and aperture magnitudes in $r$-band ($\text{mag}_r$, from the KiDS catalogue) to divide the final sample of 52 objects into four groups, from bright to very faint, 
scheduling from one to four observation blocks (OBs) of 1 hour for the different groups. More specifically, we requested one OB for the `bright' objects, those with $\text{mag}_r \le 18.0$ mag and $\langle\mu_{\mathrm {e}}\rangle_r \le 18.4$ mag/arcsec$^{2}$, two OBs for the `medium' ones ($18.0<\text{mag}_r\le19.3$ mag and $\langle\mu_{\mathrm {e}}\rangle_r\le 19.0$ mag/arcsec$^{2}$), three OBs for the  `faint' ones ($18.5<\text{mag}_r<19.3$ mag and $\langle\mu_{\mathrm {e}}\rangle_r > 19.0$ or $\text{mag}_r>19.25$ and $\langle\mu_{\mathrm {e}}\rangle_r\le 19.0$ mag/arcsec$^{2}$), and four OBs for the `very faint'  objects ($\text{mag}_r \ge 19.3$ mag and $\langle\mu_{\mathrm {e}}\rangle_r > 19.0$ mag/arcsec$^{2}$). 
In addition, we increment the final number of OBs (i.e. the resulting S/N of the spectrum) in systems where the estimated total stellar mass from photometry was \Mstar$\ge 10^{11}$M$_{\odot}$ or where an estimate of the stellar velocity dispersion was already present from the literature (e.g. from the SDSS or GAMA catalogues) and this resulted in $\sigma_{\star}>250$ km/s. This is motivated by the fact that, according to \cite{Pulsoni20}, the probability that a galaxy formed in a two-phase scenario rises for larger stellar masses ($\sim80$\% of galaxies with \Mstar$\ge 10^{11}$M$_{\odot}$). Having higher S/N on the most likely relics can be helpful to measure the IMF from gravity-sensitive features.  
The validity of this strategy was demonstrated in the \citetalias{Spiniello20_Pilot}, focusing on the first three objects for which ESO delivered complete observations. 
The distribution in $r$-band magnitude and surface brightness at the effective radii, used to split the systems into `brightness groups',  are shown in the histograms in the lower panels of  Figure~\ref{fig:obj_distrib}, with the same colour coding already described. 
By the end of 2020, 66  of the 154 hours of observations had been carried out on 24 galaxies in total.  
Of these, the 19 galaxies with complete observations (including the 3 already published in the Pilot Paper) form the \INSPIRE\, Data Release 1 presented in this paper and released to ESO as phase 3 collection\footnote{The data can be found on the ESO Archive Science Portal:  \url{http://archive.eso.org/scienceportal/home?data_collection=INSPIRE}.}. 
The \INSPIRE\, DR1 targets along with their coordinates, $r$-band magnitudes, and surface brightnesses (used to split the systems into OB groups) are listed in Table~\ref{tab:sample} along with the number of OBs scheduled for that system, the position angle (P.A.) of the slit, and the median seeing of the observations\footnote{We quote the median of the different HIERARCH ESO TEL IA FWHM keywords listed in the header of each OB.}. The first block of columns in Table~\ref{tab:sample_KIDS} lists, instead, the original paper from which the candidate is taken and the spectroscopic redshift we measure from the XSH spectra.

\begin{table*}
\centering
\begin{tabular}{lrrccccccc}
\hline
\hline
  \multicolumn{1}{c}{ID} &
  \multicolumn{1}{c}{RA} &
  \multicolumn{1}{c}{DEC} &
  \multicolumn{1}{c}{mag$_{r}$} &
  \multicolumn{1}{c}{$\langle\mu_{\mathrm {e}}\rangle_r$}&
  \multicolumn{1}{c}{OBs} &
  \multicolumn{1}{c}{Exp.Time} &
  \multicolumn{1}{c}{P.A.} &
  \multicolumn{1}{c}{$\langle\mathrm{Seeing UVB}\rangle\,$} & 
    \multicolumn{1}{c}{$\langle\mathrm{Seeing VIS}\rangle\,$} \\

  \multicolumn{1}{c}{ } &
  \multicolumn{1}{c}{(deg)} &
  \multicolumn{1}{c}{(deg)} &  \multicolumn{1}{c}{(AB)} &
  \multicolumn{1}{c}{(AB)} &
  \multicolumn{1}{c}{\#} &
  \multicolumn{1}{c}{(sec) } &
  \multicolumn{1}{c}{(deg) } &
  \multicolumn{1}{c}{(arcsec)}&   
  \multicolumn{1}{c}{(arcsec)}\\
\hline
KiDS  J0211-3155 & 32.8962202 & -31.9279437 & 19.78 & 18.38 & 4/4  & 11240 & 289.3 & 0.88 & 0.85 \\
KiDS  J0224-3143 & 36.0902655 & -31.7244923  & 19.25 & 17.98 & 4/4 & 11240 & 311.9 & 1.08   & 1.00 \\
KiDS  J0226-3158 & 36.5109217 & -31.9810149 & 19.25 & 18.83 & 3/3 & 8430 &   33.4 & 0.99  &  0.96\\
KiDS  J0240-3141 & 40.0080971 & -31.6950406 & 19.05 & 17.00 & 2/2 & 5620 &  267.7 & 0.90  & 0.89 \\
KiDS  J0314-3215 & 48.5942558 & -32.2632679  & 19.57 & 16.98 & 3/3 & 8430 &  263.4 & 0.74  & 0.71 \\
KiDS J0316-2953 & 49.1896388  & -29.8835868 & 19.66 & 18.09 & 3/3 & 8430 &   342.1 & 0.98  & 0.99 \\
KiDS  J0317-2957 & 49.4141028 & -29.9561748 & 19.1 & 18.02 & 2/2 & 5620 & 
  339.9 & 0.86  & 0.87 \\
KiDS  J0321-3213 & 50.2954390 & -32.2221290 & 19.23 & 18.28 & 4/4 & 11240 &   311.5 & 0.80 & 0.80 \\
KiDS  J0326-3303 & 51.5140585 & -33.0540443 & 19.48 & 18.89 & 4/4 & 11240 &  354.4 & 0.90  & 0.90 \\ 
KiDS  J0838+0052 & 129.530452 & 0.88238415 & 19.29 & 19.03 & 3/3 & 8430 &
  334.7 & 0.85 & 0.85 \\
KiDS  J0842+0059 & 130.666536 & 0.98993690 & 19.6 & 18.36 & 3/3 & 8430 & 
  346.8 & 0.86 &  0.85\\
KiDS  J0847+0112 & 131.911239 & 1.20571289 & 18.41 & 18.69 & 2/2 & 5620 &
  314.4 & 0.94 & 0.96\\
KiDS  J0857-0108 & 134.251219 & -1.14570771 & 19.21 & 19.02 & 2/2 & 2810 & 356.2 & 0.88 & 0.83 \\
KiDS  J0918+0122 & 139.644643 & 1.37947803 & 19.13 & 18.47 & 3/3 & 8430 &   301.0 & 0.89 & 0.89 \\
KiDS  J0920+0212 & 140.232084 & 2.21268315 & 18.87 & 18.54 & 2/2 & 5620 & 270.5 & 0.94 & 0.96 \\
KiDS J2305-3436 & 346.335663 & -34.6030907 & 19.69 & 18.93 & 4/4 & 11240 & 316.4 & 0.92 & 0.93 \\
KiDS J2312-3438 & 348.238904 & -34.6485914 & 19.32 & 18.34 & 9/9* & 25290 & 275.4 & 1.40 & 1.33 \\
KiDS J2327-3312 & 351.991016 & -33.2007599 & 19.35 & 18.28 & 4/4 & 11240 & 315.6  & 0.92 & 0.92 \\
KiDS J2359-3320 & 359.985168 & -33.3335828 & 19.59 & 18.07 & 4/4 & 2810 & 265.1 & 0.90 & 0.87 \\
\hline
\hline
\end{tabular}
    \caption{The \INSPIRE\, DR1 sample. Together with ID and coordinates, we give, for each relic candidate, the 
    aperture magnitudes (MAG\_AUTO from the KiDS DR3 catalogue) and the surface brightness luminosities averaged within the effective radius ($\langle\mu_{\mathrm {e}}\rangle$), both in $r$-band, that have been used to decide the number of OBs scheduled for each system, which is also listed in the table.  $^*$For J2312-3438 a total of four OBs were requested, however five OBs were repeated because they were classified as grade C for violating the seeing constraints. This means that for this particular system we received more data than originally expected but of lower quality. 
    Furthermore, the total exposure time on target, the P.A. to which the slit was oriented (on the major axis of the galaxies), and the median seeing during observations both in UVB and VIS arm  are also given. These values are the median of the  headers keyword HIERARCH ESO TEL IA FWHM of each OB, representing the delivered seeing corrected per airmass.}  
   \label{tab:sample}
\end{table*}

\subsection{Morpho-photometric characteristics of the \INSPIRE\, targets from KiDS}
\label{subsec:sampleKiDS}
As part of this \INSPIRE\, data release, we provide some useful photometrical information and structural parameters derived for the relic candidates in previous papers.  
In the second block of columns of Table~\ref{tab:sample_KIDS}, we provide previously estimated redshifts, stellar masses, and median \Reff (calculated using the $g$, $r$, and $i$-bands images) from \citetalias{Scognamiglio20}. These numbers have been used for the \INSPIRE\, target selection. 
Very good agreement is found between the previous and newly estimated spectroscopic  redshifts for all the objects: the differences between them are $\le0.01$  in 18/19
cases and still $<0.05$  in the remaining case (KiDS  J0316-2953).

In the third  column-block of the same table, we provide, for each UCMG, seeing-deconvolved structural parameters computed from the $r-$band KiDS image, the one with the best seeing.  Specifically, from left to right we list the effective radii in arcsec, \Reff, and kiloparsecs, the S\'ersic index, $n$, and axis ratio, $q$. 
These are calculated in  \citetalias{Tortora+18_UCMGs} and \citetalias{Scognamiglio20} fitting a PSF convolved S\'ersic profile to the images using the code \textsc{2dphot} \citep{LaBarbera_08_2DPHOT}. The effective radii are  translated into kiloparsecs using the Python version of the Ned Wright's Cosmology Calculator\footnote{\url{http://www.astro.ucla.edu/~wright/CosmoCalc.html}}.    
Uncertainties on the sizes, given the range covered by the  \INSPIRE\, objects, are of the order of 20\%-50\%, as visible from the Figure~B2 in \citetalias{Tortora+18_UCMGs}. 

\begin{figure*}
\includegraphics[width=18cm]{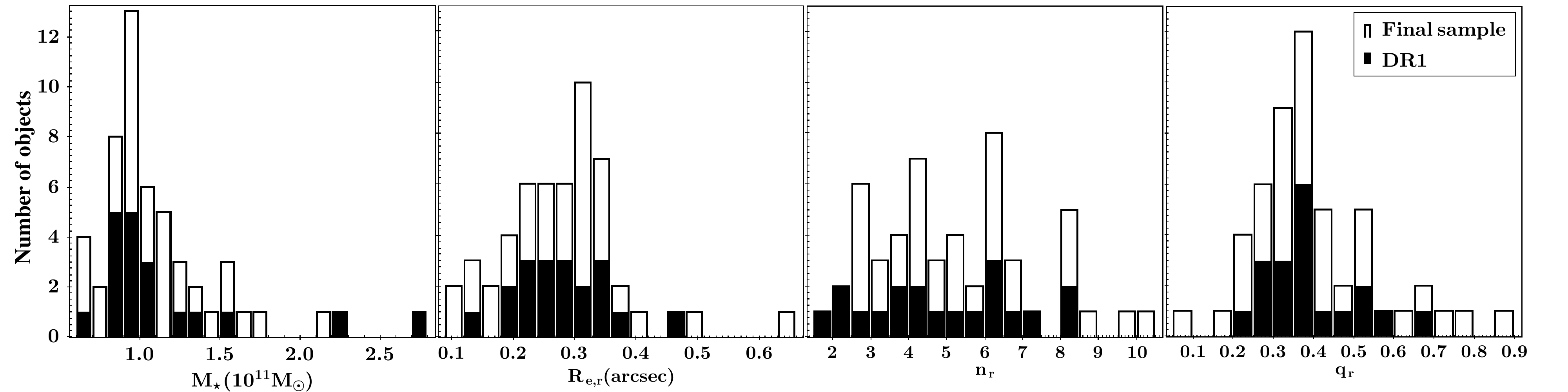}
\caption{Distribution of the stellar masses and of  the structural parameters obtained from the $r$-band KiDS images. 
From left to right we show the histograms of the total stellar masses in units of 10$^{11}$ solar masses and bins of 0.1 (calculated in \citetalias{Scognamiglio20}), $r-$band sizes in arcseconds with bins of 0.03, the S\'ersic indices (n) with bins of 0.3 and axis ratios (q) with bins of 0.05, both in $r-$band  (computed in \citealt{Roy+18}).}
\label{fig:struct_param_distrib}
\end{figure*}

The histograms showing the distribution of the $r-$band structural parameters and the optical stellar masses for the objects presented as part of this DR1 (filled) and for the final \INSPIRE\, dataset (empty) are shown in Figure~\ref{fig:struct_param_distrib}. 

\begin{table*}
\begin{tabular}{lcc|cccc|cccc|cc}
\hline
\hline
\multicolumn{1}{c}{ID} &
\multicolumn{1}{c}{Sample} &
\multicolumn{1}{c|}{z$_{\mathrm{XSH}}$} & 
\multicolumn{1}{c}{z$_{\mathrm{prev}}$} &
\multicolumn{1}{c}{M$_{\star}$} & 
\multicolumn{1}{c}{$\langle\mathrm{R}_{\mathrm{e}}\rangle\,$} &
\multicolumn{1}{c|}{$\langle\mathrm{R}_{\mathrm{e}}\rangle\,$} & 
\multicolumn{1}{c}{$\mathrm{R}_{\mathrm{e,r}}$} &
\multicolumn{1}{c}{$\mathrm{R}_{\mathrm{e,r}}$} &
\multicolumn{1}{c}{$n_{\mathrm{r}}$} &
\multicolumn{1}{c|}{$q_{\mathrm{r}}$} & 
\multicolumn{1}{c}{R50$_{\mathrm{UVB}}$} & 
\multicolumn{1}{c}{R50$_{\mathrm{VIS}}$} \\
\multicolumn{1}{c}{KiDS} &
\multicolumn{1}{c}{ } &
\multicolumn{1}{c|}{ } &
\multicolumn{1}{c}{ } &
\multicolumn{1}{c}{($10^{11}$M$_{\odot}$)}  &
\multicolumn{1}{c}{($\arcsec$)} &
\multicolumn{1}{c|}{(kpc)} & 
\multicolumn{1}{c}{($\arcsec$)} &
\multicolumn{1}{c}{(kpc)} & 
\multicolumn{1}{c}{ } &
\multicolumn{1}{c|}{ } &
\multicolumn{1}{c}{($\arcsec$)} &
\multicolumn{1}{c}{($\arcsec$)} \\
\hline
J0211-3155     & T18  & 0.3012 & 0.30135 & 0.876  & 0.237 & 1.07 &  0.237 &    1.07  & 8.1 & 0.48 &  0.58 & 0.55  \\
J0224-3143     & T18  & 0.3839 & 0.3853  & 2.710  & 0.292 & 1.55 &  0.247 &    1.31  & 6.1 & 0.39 &  0.55 & 0.55 \\
J0226-3158     & T18  & 0.2355 & 0.2364  & 0.688  & 0.350 & 1.32 &  0.337 &    1.27  & 3.7 & 0.60 &  0.58 & 0.50 \\
J0240-3141     & T18  & 0.2789 & 0.2809  & 0.981  & 0.189 & 0.81 &  0.189 &    0.81  & 8.1 & 0.27 &  0.60 & 0.50 \\
J0314-3215     & T18  & 0.2874 & 0.2888  & 1.000  & 0.150 & 0.66 &  0.148 &    0.65  & 5.5 & 0.39 &  0.55 & 0.50  \\
J0316-2953     & T18  & 0.3596 & 0.3102  & 0.873  & 0.200 & 1.02 &  0.200 &    1.02  & 3.5 & 0.31 &  0.58 & 0.50 \\
J0317-2957     & T18  & 0.2611 & 0.2610  & 0.873  & 0.257 & 1.05 &  0.257 &    1.05  & 5.0 & 0.21 &  0.59 & 0.50 \\
J0321-3213     & T18  & 0.2947 & 0.2953  & 1.233  & 0.309 & 1.37 &  0.281 &    1.25  & 4.9 & 0.39 &  0.60 & 0.50 \\
J0326-3303     & T18  & 0.2971 & 0.2976  & 0.930  & 0.321 & 1.44 &  0.321 &    1.44  & 3.7 & 0.35 &  0.60 & 0.50 \\
J0838+0052     & S20  & 0.2702 & 0.2703  & 0.870  & 0.306 & 1.28 &  0.351 &    1.47  & 4.0 & 0.41 &  0.55 & 0.50 \\
J0842+0059     & S20  & 0.2959 & 0.2946  & 0.910  & 0.226 & 1.01 &  0.226 &    1.01  & 3.3 & 0.29 &  0.55 & 0.50 \\
J0847+0112$^*$ & T18  & 0.1764 & 0.17636 & 0.989  & 0.456 & 1.37 &  0.465 &    1.40  & 3.3 & 0.26 &  0.60 & 0.55 \\
J0857-0108     & S20  & 0.2694 & 0.2696  & 1.000  & 0.336 & 1.40 &  0.366 &    1.53  & 2.9 & 0.33 &  0.60 & 0.55 \\
J0918+0122     & T18  & 0.3731 & 0.3733  & 2.263  & 0.328 & 1.71 &  0.328 &    1.71  & 6.0 & 0.51 &  0.60 & 0.55 \\
J0920+0212$^*$ & T18  & 0.2800 & 0.2797  & 1.031  & 0.345 & 1.48 &  0.345 &    1.48  & 2.0 & 0.32 &  0.58 & 0.55 \\
J2305-3436     & T18  & 0.2978 & 0.30146 & 0.857  & 0.307 & 1.29 &  0.293 &    1.23  & 3.4 & 0.40 &  0.55 & 0.50 \\
J2312-3438     & T18  & 0.3665 & 0.3664  & 1.344  & 0.243 & 1.25 &  0.243 &    1.25  & 2.3 & 0.43 &  0.60 & 0.60 \\
J2327-3312     & T18  & 0.4065 & 0.40661 & 1.569  & 0.276 & 1.15 &  0.276 &    1.51  & 5.9 & 0.67 &  0.58 & 0.50  \\
J2359-3320     & T18  & 0.2885 & 0.28922 & 1.065  & 0.237 & 1.03 &  0.237 &    1.04  & 4.4 & 0.39 &  0.58 & 0.50 \\
\hline
\hline
\end{tabular}
\begin{flushright}
\footnotesize{$^*$Also observed in GAMA.}
\end{flushright}
\caption{Spectroscopic, sample-selection, and structural characteristics of the \INSPIRE\, DR1 sample. In the first block of columns, we provide the KiDS publication from where each candidate was taken and the new redshifts inferred from XSH spectra. The first column of the second block lists instead the previously estimated redshifts. The two values are always in fair agreement. The previous redshifts, as well as the median ($gri$) size and stellar masses also provided in the same block, have been used for the INSPIRE object selection. The third block lists the structural parameters presented in \citetalias{Tortora+18_UCMGs} and/or \citetalias{Scognamiglio20} and obtained from the $r$-band KiDS images (the image with the best seeing), which are also shown in the histograms in Figure~\ref{fig:struct_param_distrib}, together with stellar masses. Finally, in the last block of columns, we report the apertures that encapsulate 50\% of the light for the UVB and VIS arm, respectively (see Sect.~\ref{subsec:1d-extr}). }
\label{tab:sample_KIDS}
\end{table*}

\section{Data reduction}
\label{sec:data_analysis}
ESO releases only the 1D spectra as Internal Data Products (IDPs), but we need to customise the extraction (see following section). Therefore, we reduced all the data ourselves using the ESO XSH pipeline (v3.5.0) under the ESOReflex Workflow (\citealt{Freudling+13}, version 2.11) up to the creation of the two-dimentional (2D) spectral frames. We have however tested the data reduction results we obtained against that obtained directly from ESO as part of the IDPs production finding 2D spectra of  similar quality. 
For the VIS arm, we also corrected the spectra for the presence of telluric lines using the code molecfit (\citealt{Smette15}, version 3.0.3) run with its interactive ESOReflex workflow. 
The extraction of the 1D spectra, and the combination of the multiple OBs is instead carried out using the same IDL and Python routines already employed in the \citetalias{Spiniello20_Pilot}.  

In this DR1, we limit our analysis to the UVB and VIS arm, as already done in the \citetalias{Spiniello20_Pilot}, because this is enough to precisely estimate the line-of-sight velocity distribution (LOSVD) and to constrain the age, metallicity, and Mg-abundance of the stellar populations, and thus obtain confirmation of the relic nature.  
In turn, the NIR will become crucial in order to infer the low-mass end of the IMF, which will be the topic of a separate forthcoming publication accompanied by the ESO data release of the NIR spectra. Finally, we also implement a sigma clipping routine to further clean the 2D spectra from cosmic ray and sky residuals. 

\subsection{Extraction of 1D spectra}
\label{subsec:1d-extr}
As illustrated by \citetalias{Spiniello20_Pilot}, the spatial resolution of XSH is not sufficient to resolve ultra compact objects such as those targeted here.  In all cases, the effective radii in arcseconds (see Table~\ref{tab:sample_KIDS}) are much smaller than the median seeing of the observations (which varied from $0.71\arcsec$ to $1.23\arcsec$; see Table~\ref{tab:sample}). 
This means that the spectra are completely seeing dominated.
Thus, the most meaningful approach to compare the \INSPIRE\, galaxies with each other and with other samples of galaxies is to extract the spectra from an aperture that contains more or less the same fraction of light for the different objects, given the seeing during observations and their surface brightness profiles. This is the approach we followed in the Pilot Study: we extracted the surface brightness profiles directly from the 2D spectra of each OB\footnote{Collapsing the frames in the dispersion direction and assuming spherical symmetry and thus only considering one side of the slit.} and integrated the profiles up to when the flux reaches the null value, thus computing the `total light'. We then repeated the same integration but stopped at different radii and finally took the ratio between these aperture fluxes and the total to obtain the fraction of light within each extracted aperture. Differently from the previous case, where we considered an aperture encapsulating $\sim30$\% of the light, in this paper we took a slightly larger radius, summing up to $\sim50$\% of the total light\footnote{We consider the median aperture radius among the different OBs.}. Generally, this roughly corresponds to extracting the spectra from approximately $6$-$8$ pixels (1 pixel=$0.16\arcsec$). We obtain in this way an integrated 1D spectrum comparable in light fraction to that extracted at one effective radius, \Reff, but taking into account the seeing convolution. However, as we demonstrate in Appendix~\ref{app:seeing_correction}, these values have to be interpreted as lower limits when compared to the values one would measure for resolved objects. This is because the  R50 spectrum encapsulates half of the observed light, but because of the seeing, the light gets pushed to larger radii. Hence, it is not surprising that the R50 apertures are larger than the \Reff inferred from KiDS.   Assuming that the velocity dispersion decreases with increasing distance from the centre, measuring it from the R50 aperture, which contains a mix of light from inside and outside the real effective radius, we underestimate it by a factor of $1.07$ (on average; see Appendix~\ref{app:seeing_correction} for more details).  
The R50 extraction might nevertheless come in handy when comparing the \INSPIRE\, sample with other galaxy samples from the literature, as this is the most comparable aperture ---at least in terms of light fraction--- to that extracted at one effective radius for normal sized galaxies. The aperture radii encapsulating 50\% of the light (R50, hereafter) for the UVB and VIS spectra of each system are listed in the last two columns of Table~\ref{tab:sample_KIDS}. 

On the other hand, in order to prove that the confirmed relics have assembled the great majority of stars through a single star formation episode, it is advisable to collect light from as wide a spatial region  as possible and ideally from the whole galaxy. This would allow us to exclude the presence of secondary stellar populations with different ages and metallicity 
contributing more than a few percent of the light \citep{Salvador-Rusinol2021_nature}.
However, in turn, it would also cause a considerable decrease in the final S/N, making it harder to obtain precise estimates of the stellar population parameters. To overcome this problem, here we  also follow the optimal extraction approach described in \citet{Naylor+98}. 
Thus, for each galaxy (and in each arm separately), we obtain 
and release two different 1D spectra via the ESO Phase 3 Archive Science Portal: one extracted within an aperture that includes 50\% of the light (listed in Table~\ref{tab:sample_KIDS} for each system in each arm) and one obtained with the optimal extraction (OptEx, hereafter). 
As expected, the optimal extraction procedure increases the final S/N (theoretically, an increment of $\sim10$\% is expected; see Eq. 9-11 in \citealt{Naylor+98}) with respect to the fixed aperture strategy, as shown in Appendix~\ref{app:signal_to_noise}. 

\subsection{Emission lines}
In 6 out of 19 systems we found convincing evidence for emission lines from  [OII] ($\lambda\sim3727$\AA) and [NII] ($\lambda\sim6583$\AA), including one of the confirmed relics from the Pilot Study, J0847+0112.  These are the only strong emission  lines visible in the entire UVB+VIS spectra. However, we note that weak emissions from Balmer lines arise after subtracting the much stronger stellar contribution in absorption.

As reported in \citet{Yan06}, many red, elliptical local galaxies have non-negligible [OII] and/or [NII] emission in their spectra (see also \citealt{Caldwell84, Phillips86,Gallazzi+14,Wu14b,LaBarbera+19}) but this does not necessarily indicate recent star formation. Indeed, active galactic nuclei (AGNs), and especially  low-ionization nuclear emission-line regions (LINERs, e.g. \citealt{Halpern83, Groves04, Dopita96}), fast shock waves, cooling flows, and  post asymptotic giant branch  (post-AGB, e.g. \citealt{Binette94, Yan12, Belfiore16}) stars might also produce such emission. 

Given the flux ratios [OII]/H$\beta$ and [NII]/H$\alpha$ (both always $\ge1$) we measure in the six \INSPIRE\, objects with emission lines, and in line with the literature on red and dead massive ETGs (e.g. \citealt{LaBarbera+19}), we can exclude that these lines are caused by recent star formation.   
Looking at Figure 11 in \citet{Yan06}, star forming galaxies all have $\log$([NII]/H$\alpha)<-0.2$, which would imply an H$\alpha$ emission that is twice as strong as that of [NII]. This is not what we observe for the six \INSPIRE\, objects showing emission lines, as [NII] is always as strong as the H$\alpha$ or even stronger ($0<\log$([NII]/H$\alpha)<0.5$). 

As we do not detect any [OIII]($\lambda5007$\AA), we can also rule out the Seyfert case. However, the uncertainties on the flux ratios we would obtain given the S/N of the spectra and the superposition of emission and absorption in H$\beta$ and H$\alpha$ will not allow us to further disentangle the ionising process producing these lines. Furthermore, given the line strength of the [OII], we compute a star formation rate (SFR) of slower than $0.2M_{\odot}/yr$ for all the objects\footnote{SFR $<0.01M_{\odot}/yr$ for the relics} using the method presented in \citet{Kewley04}.   Finally, we note that we find emission in relics and non-relics, and therefore we exclude a correlation between emission lines and relic nature. 
We plan to further investigate  
this matter and will publish our findings in a forthcoming publication of the INSPIRE series. 

\begin{figure}
\includegraphics[width=8.3cm]{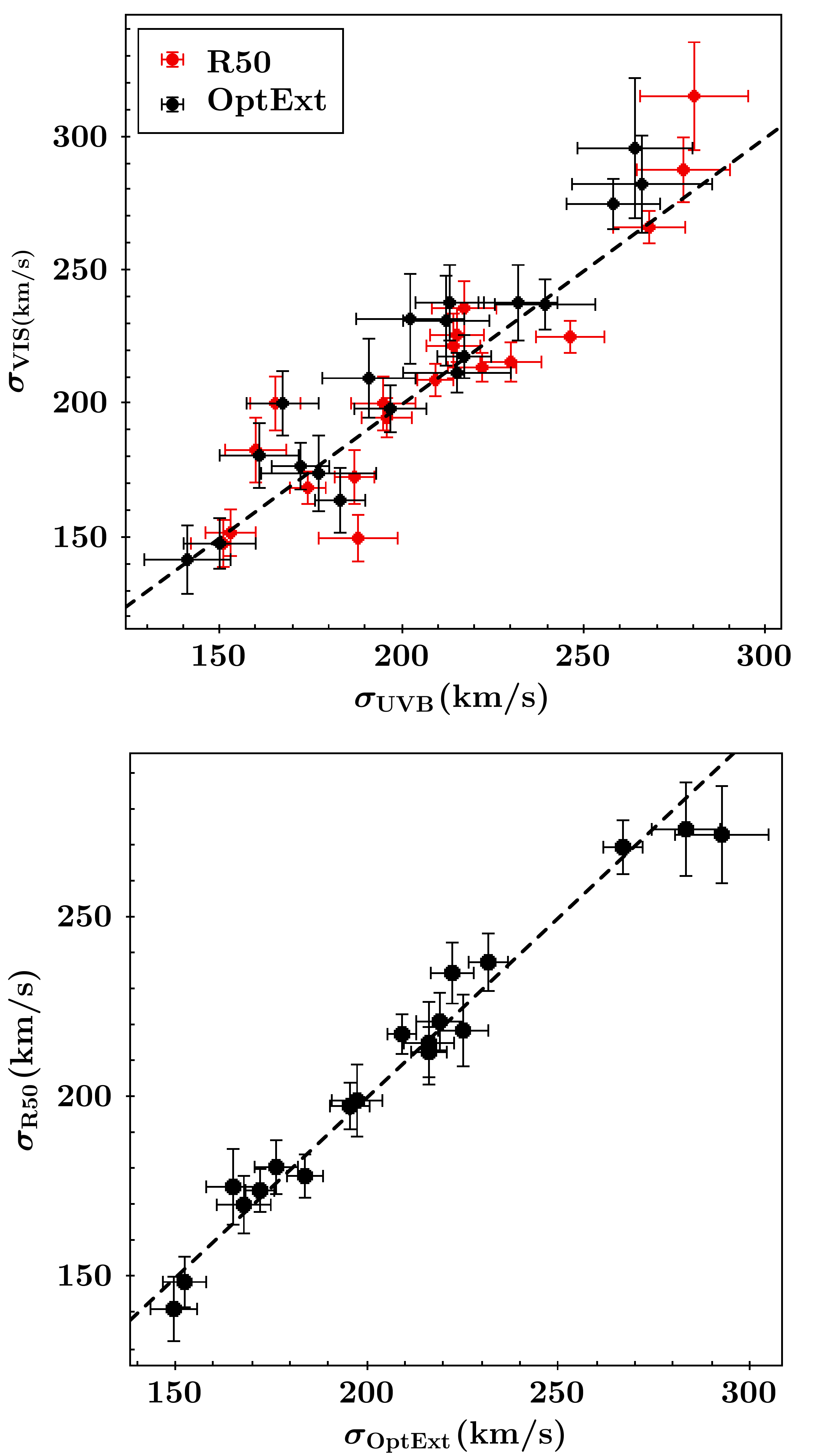}
\caption{\textit{Upper panels:} Velocity dispersion values obtained from the UVB plotted against the ones obtained from the VIS spectra extracted with the OptExt method (black points) and within the R50 apertures (red points). 
\textit{Bottom panel:} Velocity dispersion values obtained as weighted mean of the estimates in UVB and VIS for the OptExt spectra versus those obtained for the R50 ones.}
\label{fig:kinematics}
\end{figure}
 
\section{Line-of-sight velocity distribution analysis}
\label{sec:kinematics}
Another difference with respect to the \citetalias{Spiniello20_Pilot}  is that here we run the kinematics separately on the two XSH arms. 
This has two positive outcomes: it allows us to obtain two independent estimates of the velocity dispersion for each object without having to degrade the resolution of the VIS spectrum to match that of the UVB. 

To derive the LOSVD, we use the Penalised Pixel-fitting software (\ppxf; \citealt{Cappellari04,Cappellari17}) in its Python implementation (v7.4.3)\footnote{The package is available here: \href{https://pypi.org/project/ppxf/}{https://pypi.org/project/ppxf/}.}. Following the recommendations suggested by the author of the code, we use an additive Legendre polynomial (ADEGREE) to correct the continuum shape during the kinematical fit. We fix ADEGREE=20, which is justified by the test presented in the Appendix of the Pilot Paper. We tested values from 4 to 29 and found that this value stabilises the results against changes of other parameters while at the same time minimising the reduced $\chi'^{2}$ which is  equal to the $\chi^{2}$ divided by the number of good pixels used for the fit. 
As stellar templates, we use the MILES single stellar population (SSP) models by \citet{Vazdekis15}, although we have demonstrated in the Pilot Program that the kinematic results obtained using stars are fully consistent.

We run \ppxf\, on both the OptExt and R50 spectra, in the UVB and VIS separately, and without any linear regularisation to the weights of the fitting templates (i.e. we set the corresponding keyword in \ppxf\, REGUL=0). We limited the fit to the first two moments of the LOSVD, the velocity, and the velocity dispersion. Although the S/N might be just enough to 
also derive a reliable estimate for the higher moments, we note that a single integrated measure for the h$3$ and h$4$ Gauss-Hermite parameters, representing a parametrisation of   skewness and kurtosis, respectively \citep{vanderMarel93},  does not provide any relevant physical information about the causes for a non-Gaussian velocity distribution.  
For the UVB, we start the fit at $\lambda=3550$\AA, the bluest wavelength covered by the stellar templates, and go as red as possible for each galaxy, given its redshift (between 4000 and 4300\AA). The strongest stellar absorption features in the UVB are the Calcium K and H lines. 
For the VIS, we use instead the same spectral region for all galaxies ([4800-6500]\AA), covering H$\beta$, Mg, and Fe lines as well as the NaD strong absorption feature.  
In both arms, we use the function `determine\_goodpixels' provided within the \ppxf\, package to mask gas emission lines and residuals of sky subtraction  from the fit. 

The stellar velocity dispersion values obtained from the two arms agree fairly well for all the \INSPIRE\, DR1 objects, as shown in the upper panel of Figure~\ref{fig:kinematics}, for both the OptExt (black) and the R50 (red) extractions.  For the OptExt  case, a small systematic offset of $\sim10$ km/s is detected, with VIS values larger than the UVB ones. 

In the bottom panel of the same figure, we plot instead a direct comparison between the final velocity dispersion values obtained from the OptExt (x-axis) and R50 (y-axis) extractions. These values are computed as the mean of the measurements obtained from the UVB and those obtained from the VIS 
spectra (weighted by their errors) for each aperture. Good agreement is found (within $1\sigma$ errors in 18/19 objects and  $1.5\sigma$ in the remaining one) for the OptExt versus R50 values. 

All the measured stellar velocity dispersion values are  listed in Table~\ref{tab:sigmas}. The values computed in the two arms separately are listed in the second block of columns, while the  mean of the UVB and VIS values for each extraction method are reported in the third block of columns.

In Appendix~\ref{app:seeing_correction} we show that if we want to compare the R50 velocity dispersion measurements with values obtained at the \Reff for resolved objects, we need to take into account the effect of the seeing, which overall causes slight underestimation of the measurements. 
Finally, a direct comparison between the final velocity dispersion values we obtained and the ones reported in \citetalias{Scognamiglio20} is presented in Appendix~\ref{app:s20compa}. In this case, to mimic the procedure followed by \citetalias{Scognamiglio20}, we correct the R50 values to the effective radii using Eq.~1 of \citet{Cappellari+06}.  However, we note that this correction is not  appropriate in the case of UCMGs at $z>0.1$, because the objects are not resolved  (i.e. they are much smaller in size than the seeing and the slit width).

\begin{table*}
\centering
\begin{tabular}{c|llll|ll}
\hline
\hline
\multicolumn{1}{c|}{ID KiDS} &
\multicolumn{1}{c}{$\sigma_{\mathrm{OptEx, UVB}}$} &
\multicolumn{1}{c}{$\sigma_{\mathrm{R50, UVB}}$}    &
\multicolumn{1}{c}{$\sigma_{\mathrm{OptEx, VIS}}$} &
\multicolumn{1}{c|}{$\sigma_{\mathrm{R50, VIS}}$}    &
\multicolumn{1}{c}{$<\sigma_{\mathrm{OptEx}}>$} &
\multicolumn{1}{c}{$<\sigma_{\mathrm{R50}}>$}   \\
\multicolumn{1}{c|}{} &
\multicolumn{1}{c}{(km/s)} &
\multicolumn{1}{c}{(km/s)} &
\multicolumn{1}{c}{(km/s)} &
\multicolumn{1}{c|}{(km/s)} &
\multicolumn{1}{c}{(km/s)} &
\multicolumn{1}{c}{(km/s)} \\
\hline
J0211-3155 &   $212\pm12$ &  $230\pm9$  & $231\pm17$ & $249\pm10$  & $225\pm7$  & $218\pm10$ \\
J0224-3143 &   $258\pm13$ &  $282\pm10$ & $275\pm9 $ & $280\pm6 $  & $267\pm5$  & $269\pm8$  \\
J0226-3158 &   $183\pm7 $ &  $194\pm5$  & $164\pm12$ & $178\pm10$  & $184\pm5$  & $178\pm6$  \\
J0240-3141 &   $191\pm13$ &  $210\pm9$  & $210\pm15$ & $213\pm10$  & $197\pm7$  & $199\pm10$ \\
J0314-3215 &   $161\pm11$ &  $174\pm9$  & $181\pm12$ & $198\pm12$  & $168\pm7$  & $170\pm8$  \\
J0316-2953 &   $177\pm16$ &  $202\pm11$ & $174\pm14$ & $159\pm9 $  & $165\pm7$  & $175\pm10$ \\
J0317-2957 &   $167\pm10$ &  $174\pm7$  & $200\pm12$ & $209\pm10$  & $176\pm6$  & $180\pm8$  \\
J0321-3213 &   $197\pm10$ &  $206\pm7$  & $198\pm9 $ & $202\pm8 $  & $195\pm5$  & $198\pm7$  \\
J0326-3303 &   $150\pm10$ &  $159\pm7$  & $148\pm9 $ & $157\pm9 $  & $152\pm5$  & $149\pm7$  \\ 
J0838+0052 &   $172\pm8 $ &  $179\pm5$  & $177\pm9 $ & $173\pm6 $  & $172\pm4$  & $174\pm6$  \\
J0842+0059 &   $264\pm16$ &  $297\pm15$ & $296\pm26$ & $331\pm20$  & $293\pm12$ & $273\pm14$ \\
J0847+0112 &   $213\pm9 $ &  $219\pm7$  & $238\pm14$ & $229\pm10$  & $219\pm6$  & $221\pm8$  \\
J0857-0108 &   $141\pm12$ &  $156\pm9$  & $142\pm13$ & $152\pm9$   & $149\pm6$  & $141\pm9$  \\
J0918+0122 &   $239\pm14$ &  $256\pm9$  & $237\pm10$ & $233\pm6$   & $232\pm5$  & $238\pm8$  \\
J0920+0212 &   $232\pm11$ &  $238\pm8$  & $238\pm14$ & $223\pm7$   & $222\pm5$  & $234\pm9$  \\
J2305-3436 &   $266\pm19$ &  $289\pm13$ & $282\pm18$ & $298\pm12$  & $283\pm9$  & $274\pm13$ \\
J2312-3438 &   $217\pm7 $ &  $222\pm5$  & $218\pm8 $ & $222\pm6$   & $209\pm4$  & $217\pm6$  \\
J2327-3312 &   $215\pm15$ &  $233\pm9$  & $212\pm8 $ & $223\pm5$   & $216\pm5$  & $213\pm7$  \\
J2359-3320 &   $202\pm15$ &  $227\pm8$  & $232\pm17$ & $233\pm12$  & $216\pm6$  & $215\pm11$ \\
\hline
\hline
\end{tabular}
\caption{Velocity dispersion measurements. The second block of columns lists the values computed with the two extraction methods, from the two arms separately, and the R50 values corrected to the effective radii. The third block of columns lists instead the final values obtained as mean of the values from the two arms, for each aperture. }
\label{tab:sigmas}
\end{table*}

On the right-hand side  of Figure~\ref{fig:spec_example}, 
we show examples of the kinematics fits (top-right panels) in the UVB and VIS arms for the systems with the 
spectra with the best and with the worst mean spectral S/N. 
For each of the 19 \INSPIRE\, DR1  systems, we obtain a total of eight different estimates of the S/N, as described in Appendix~\ref{app:signal_to_noise}. The mean values of these estimates range between 55 (J2312-3438, top panel) and 22 per \AA\ (J0842+0059, bottom panel) for all the \INSPIRE\ DR1 objects. 

In the same figure, on the left, we show the $1\arcmin \times 1\arcmin$ cut out of the KiDS $r$-band image as well as a $gri$-colour zoomed-in panel of $20\arcsec \times 20\arcsec$  centred on the relic candidate. The galaxies show a variety of morphologies, ranging from disky to spheroidal,  
hence confirming the broad distribution in S\'ersic indices and axis ratios (see Fig.~\ref{fig:struct_param_distrib}). A few objects seem to be relatively isolated while some other images might suggest a group or cluster environment. In a forthcoming publication of the \INSPIRE\, series, we plan to characterise the local environment in which relics live,  
searching for possible correlations with structural, photometric or stellar population parameters (see e.g. \citealt{Ferre-Mateu+17}).  

\section{Stellar population analysis}
\label{sec:stelpop}
To constrain the stellar population parameters with \ppxf\, it is instead necessary to join together the UVB and VIS arms. This is crucial in order to have a final spectrum covering a wavelength range that is large enough to break the age--metallicity \citep{Worthey+94} degeneracy and thus properly infer the stellar population parameters.  
To perform the combination, we first brought  the UVB and VIS spectra to the same final resolution\footnote{The XSH nominal resolutions are: $\mathrm{R}_{\mathrm{UVB}}=3200$ and $\mathrm{R}_{\mathrm{VIS}}=5000$ in the UVB and VIS respectively.}. 
We used the spectral convolution procedure already introduced by \citetalias{Spiniello20_Pilot}, and we refer the reader to this latter paper for a more detailed explanation. 
The final full width at half maximum of the spectrum is chosen to be equal to that of the MILES single stellar population (SSP) models \citep{Vazdekis15} that we use in the stellar population analysis: FWHM$_{\mathrm{fin}} = 2.51$\AA\, at the rest frame wavelength. 
In particular, for the fit, we select MILES SSPs with a bimodal stellar IMF with a fixed slope of $\Gamma = 1.3$ and BaSTI theoretical isochrones\footnote{
\href{http://www.oa-teramo.inaf.it/BASTI}{http://www.oa-teramo.inaf.it/BASTI}.} with ages ranging from 1 Gyr to the age of the Universe at the redshift of each object\footnote{Given the sampling of the SSPs in age ($\Delta_{\text{age}} = 0.5$ Gyr), we choose, for each galaxy, the model with an age as close as possible to the age of the Universe at that redshift.}, with steps of $\Delta t = 0.5$ Gyr, and five 
different total metallicity bins (in terms of the total stellar metal abundance,  [M/H] = $\{-0.66,-0.25,+0.06,+0.26,+0.44\}$). 
In addition, we linearly interpolate the models in the [$\alpha/$Fe] space between the two publicly available abundances of 0.0 (solar) and 0.4, building a grid with $\Delta=0.1$. 

As the original version of the \ppxf\, code only works in the 2D parameter
space defined by age and metallicity, we first estimate the [Mg/Fe] (a proxy for [$\alpha$/Fe], which is the parameter we can change in the SSP models) via line-index strengths of the Mg$_b$ 5177 and of all the iron (Fe) lines between 3500 and 7000 \AA. 
As in the \citetalias{Spiniello20_Pilot}, we use the code SPINDEX from \citet{Trager08} to calculate the line-index strengths of the Mg$_b$ and the many iron (Fe) lines present in the wavelength range at our disposal. We then produce  the classical Mg$_b$-Fe index--index plot shown in Figure~\ref{fig:alphafe}, which allows us to obtain an estimate of the [$\alpha$/Fe]. In this index--index plot, while age and metallicity are degenerate and would move the object from the bottom left corner to the top right one, the [$\alpha$/Fe] varies in an orthogonal direction. We overplot on the galaxy indices MILES SSP models with age ranging from 5 to 10 Gyr and metallicities ranging from -0.66 to +0.44 and colour code them according to their [$\alpha$/Fe], interpolating the two available models with  [$\alpha$/Fe]=0.0 (blue) and [$\alpha$/Fe]=0.4 (yellow) in steps of $\Delta$[$\alpha$/Fe]=0.1. Without making any assumption about the stellar age and metallicity, one can directly estimate the [$\alpha$/Fe], taking the closest model (colour) to each point.  We then assigned an error bar to the measured [Mg/Fe] that is equal to the step between the models ($0.1$).

\begin{figure*}
\centering
\includegraphics[width=17.5cm]{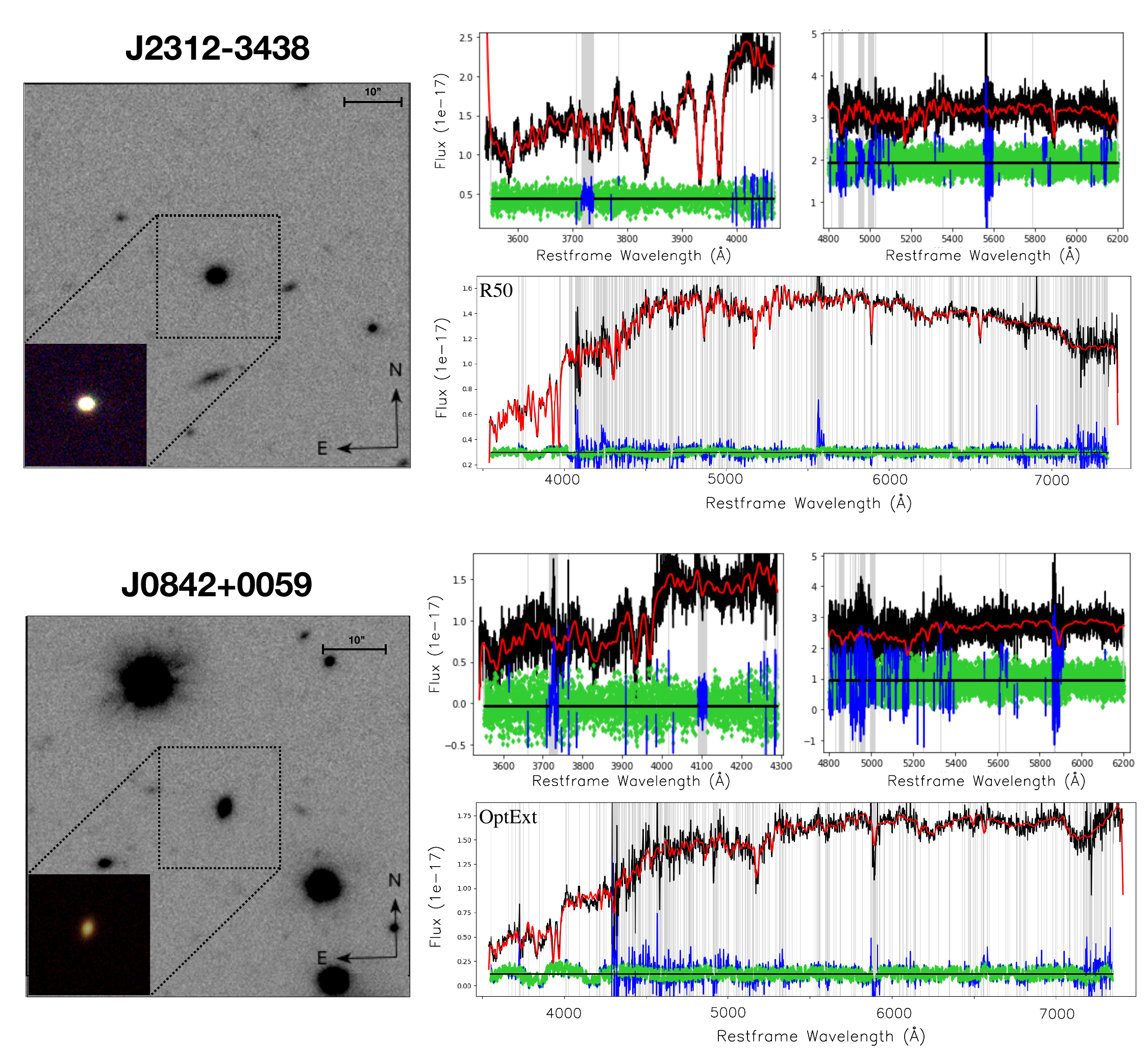}
\caption{Two of the 19  target systems of this \INSPIRE\, DR1. From top to bottom: J2312-3438, the object with the spectrum with the highest mean S/N ($\sim55$ per \AA), and J0842+0059, that with  the lowest mean spectral S/N ($\sim22$ per \AA; see Table~\ref{tab:snr_calculation} in Appendix~\ref{app:signal_to_noise}). For each object, the \textit{left panel} shows a $1\arcmin \times 1\arcmin$ cutout of the KiDS $r$-band image with the relic candidate in the middle. The \textit{zoom-in panel} shows a $gri$-colour (KiDS) image of $20\arcsec \times 20\arcsec$. The \textit{right top panels} show the result from the \ppxf\, kinematic fit (left UVB, right VIS, described in Sect.~\ref{sec:kinematics}), while the \textit{right bottom panel} shows the stellar population fitting (UVB+VIS, Sect.~\ref{sec:stelpop}) one time in the case of the R50 spectrum (top) and one time for the OptExt (bottom). In all the panels, we plot in red the best-fit stellar template, in green the residuals and in blue and grey bands the pixels which have been masked out from the fit. All the spectra are plotted at the restframe wavelength and in units of Flux normalised by $10^{-17}$. In the  kinematic fit the black spectra are plotted at their original resolution. In the stellar population one, these are smoothed to a resolution of FWHM$ = 2.51$\AA.  }
\label{fig:spec_example}
\end{figure*}

Subsequently, we ran the full spectral fitting code using models with the measured value of [$\alpha$/Fe]$_{\text{measured}}$ as well as with models with [$\alpha$/Fe]= $\pm 0.1 + [\alpha$/Fe]$_{\text{measured}}$, to take into account the uncertainties associated with the inference from line indices that are assumed to be equal to the step between two models ($0.1$). 
We stress that the estimates we obtain must be interpreted as tentative, as we do not perform any quantitative likelihood-based analysis combining a large numbers of indices, which is the only way to break the degeneracies between the different stellar population parameters (e.g. \citealt{Spiniello+14}). 
However, we anticipate that this does not play any role in the relic confirmation (see Sect.~\ref{sec:results_confirmation} for more details). 

The [Mg/Fe] values are listed in the last column of Table~\ref{tab:stelpop} and are almost always super-solar. 
Furthermore, the estimates obtained from the OptExt spectra and those obtained from R50 are generally in agreement within the errors. Only for J0226-3303, the [Mg/Fe] inferred in the OptExt case is larger ($0.2\pm0.1$) than that obtained for the R50 spectrum ($0.0\pm0.1$). We visually inspected the two spectra and checked that this was not caused by a contamination in the line strengths but genuinely reflects a difference in the strength of the iron lines. This most likely indicates the presence of spatial gradients in the $\alpha$-abundance of the stellar population of this galaxy and suggests that this galaxy is not a relic  
(which is confirmed in Section~\ref{sec:results_confirmation}). If an object is a true relic, it should have a flat [Mg/Fe] gradient with a very high value (i.e. $\ge0.3$) because it formed its stars in a very quick high-z burst of star formation which then rapidly quenched. This would prevent iron pollution of the interstellar medium caused by Type Ia supernova explosions \citep[e.g.][]{Gallazzi+06, Gallazzi21,Thomas+05}. 
Super-solar metallicities and generally old ages for the objects in the INSPIRE sample are expected too according to the mass--metallicity relation \citep[e.g.][]{Gallazzi+05, Gallazzi21} and given the colour selection presented in Section~\ref{sec:survey_presentation}. 

We ran \ppxf\, in the exact same configuration for the spectra extracted with the two different methods, deriving two independent yet not physically identical estimates of the mass-weighted stellar age and metallicity for each galaxy. 
As already done in the pilot, we performed the fit using the two extreme values of the regularisation parameter, thus deriving the minimally (REGUL = 0) and maximally (MAX\_REGUL) smoothed solutions consistent with each  spectrum and its S/N. 
Furthermore, for the stellar population run of \ppxf, we substituted the additive polynomial (set to ADEGREE=-1) ---which could affect the line strength of the spectral features--- with a multiplicative one (MDEGREE=10; see the tests performed in Appendix A of the  \citetalias{Spiniello20_Pilot}). 
Finally, as starting guesses for velocity and velocity dispersion, we used the values obtained in Section~\ref{sec:kinematics} (listed in Table~\ref{tab:sigmas}), allowing for a variation of $3\sigma$. 
We use the rest frame spectral range from $3500$ to $7000$ \AA .

To estimate the uncertainties on age and metallicity associated with the spectral noise, in the assumption of Gaussian independent errors on each pixel, we followed the `Monte Carlo perturbation' prescriptions given in the  \citetalias{Spiniello20_Pilot}: we repeated the fit 256 times, randomly simulating  a new spectrum at each run according to the $1\sigma$ RMS noise level obtained during the first fit (i.e. we rescaled the noise by maintaining the reduced $\chi'^{2}$ between 0.9 and 1.1), also randomly  varying the MDEGREE (within $\pm 4$ from the original value) and the REGUL parameter (which is nevertheless always kept smaller than the MAX\_REGUL). 
This semi-automatic  procedure produces a distribution of ages and metallicities for each spectrum, which we use to calculate the uncertainties (as standard deviation) of the final mass-weighted age and metallicity. The central values are instead those obtained from the original observed spectrum using the best-fit configuration (minimum $\chi^{2}$, MDEGREE=10). 
Two examples of the stellar population fitting are shown in the bottom right-hand panels of  Figure~\ref{fig:spec_example}, for the spectra with the highest (top) and lowest (bottom) S/N,  together with the previously described kinematical ones. 

\begin{figure}
    \centering
    \includegraphics[width=\columnwidth]{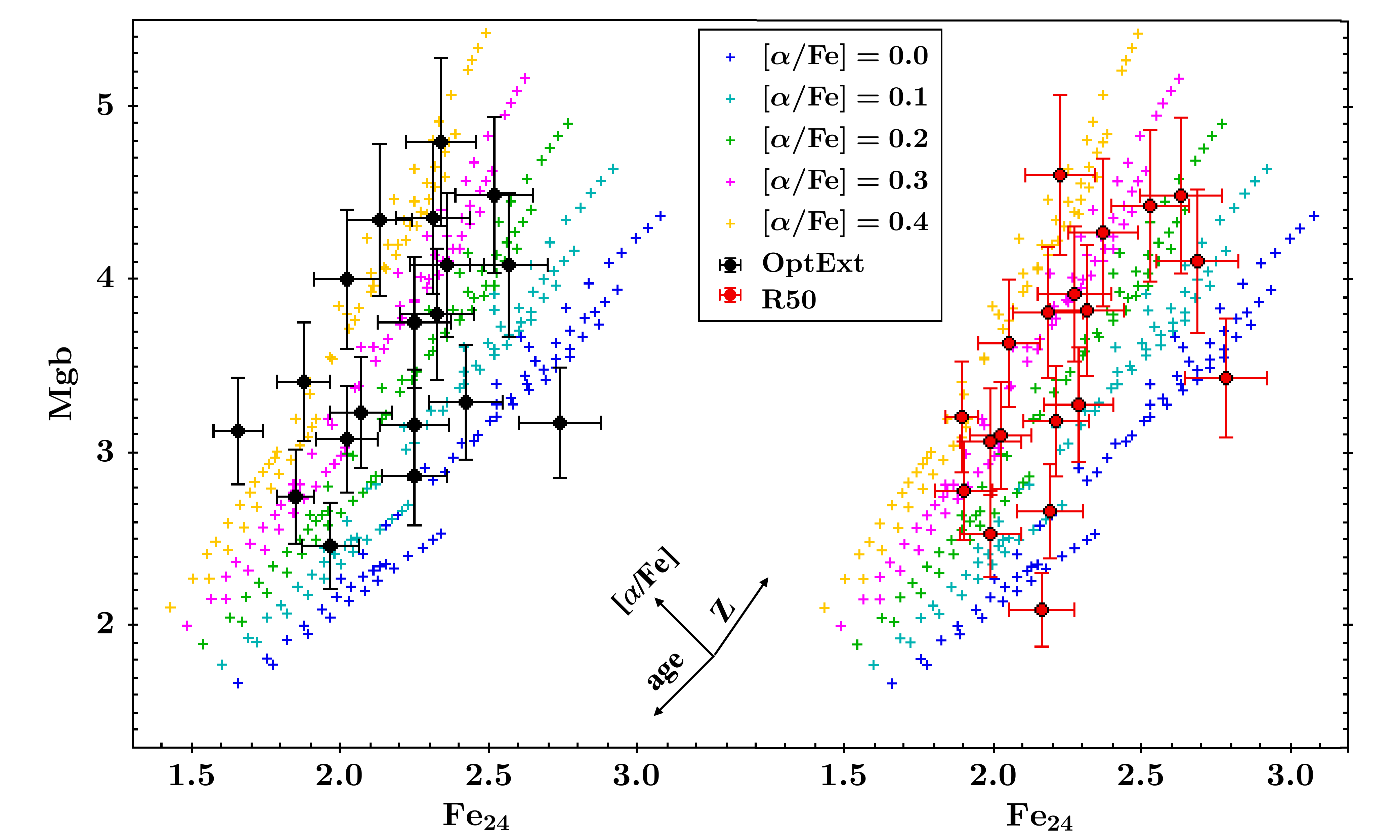}
    \caption{Mg$_b$-Fe24 index--index plot used to infer the [Mg/Fe] for the OptExt (black, left) and R50 (red, right) spectra. The F$_{24}$ index is the mean of the 24 iron lines present in the spectral range [$3500-6500$]\AA. Coloured crosses are MILES SSPs with different ages and metallicities and colour-coded according to their [$\alpha$/Fe] value. Black arrows show the direction of the variation expected by varying the stellar population parameters (age, Z, [$\alpha$/Fe]).  Both galaxies and SSP line strengths are calculated after convolving the spectra to a common resolution of $\sigma=300$ \kms. }
    \label{fig:alphafe}
\end{figure}

\section{Results}
\label{sec:results}
In this section we present the main scientific results of this paper. The first one is the derivation of the stellar population parameters of the 19 UCMGs analysed in this DR1, which allow us to confirm some of them as relics. Thanks to the larger number of systems, we can also better quantify the existence of a `degree of relicness', already hinted at in the  
\citetalias{Spiniello20_Pilot} and \citetalias{Ferre-Mateu+17}. We also compute a preliminary lower limit for the number density of relics at $0.17<z<0.39$ and highlight a possible dynamical difference between relics and non-relics. 

\subsection{Stellar populations}
The results of the stellar population analysis are summarised in the central block of columns in Table~\ref{tab:stelpop}, where we list the central ages and metallicities for the two extreme cases (REGUL = 0 and REGUL = MAX\_REGUL). 
In at least half of the systems, the MAX$\_$REGUL is small, and the results for the OptExt and R50 case are perfectly consistent, probably indicating a lack of spatial gradients (although we stress that we do not spatially resolve the systems with XSH). 
There are some galaxies for which the age difference between the REGUL = 0 and the REGUL = MAX\_REGUL is large; these are also the systems with the highest maximum regularisation and the lowest mass-weighted ages. The likely outcome is that these objects will not be confirmed as relics, as we show in the following section. 

Overall, the stars in the majority of the systems are old, often as old as the Universe. 
This demonstrates that our selection criteria are robust and that the \INSPIRE\, targets are optimal relic candidates. 
Only one system, KiDS J2327-3312, turned out to be much younger than expected from its colours ($t_{\text{gal}} \sim3-4$ Gyr). Such young age is also confirmed by the presence of very strong Balmer lines in the XSH spectra.

For the two confirmed relics already presented in the  \citetalias{Spiniello20_Pilot}, we find similar mass-weighted ages and metallicities, consistent within the errors with the previous measurements. 
For the non-relic J0314-3215, the values of [Mg/Fe] estimated from line indices are not in good agreement, with the measurement obtained in the Pilot ($0.3\pm0.1$ vs. $0.1 - 0.2$ measured here) being larger than the one presented here (for both extractions). Moreover, in the Pilot, for J0314-3215 we measure a lower limit on the integrated age of $7.9\pm0.5$ Gyr (from the MAX\_REGUL case). Here,  the integrated lower limit on the age goes down to $7.8\pm0.2$ Gyr for R50 (50\% of the light) and $6.4\pm0.3$ Gyr for OptExt (the whole galaxy but putting more weight on the pixels with higher S/N). As we use exactly the same method and tools, the only difference with respect to the Pilot Study is the extraction method. Previously, we analysed spectra extracted from a smaller aperture encapsulating only $\sim30$\% of 
the light. Here, with R50 we take $\sim50$\% of the light and with OptExt we use the whole slit, collecting light from a larger region of the galaxy. By doing so, we incorporate the contribution of more external regions  into the final 1D spectra, which is where the accreted population (if existing) would predominantly sit. This is exactly the case for the non-relic object of the \citetalias{Spiniello20_Pilot} paper. 
Hence, 
although the biggest limitation of \INSPIRE\ is that we do not resolve the objects, we demonstrated that, from  seeing-dominated spectra at different apertures, we can still detect accreted populations, if present. 

\begin{figure}
\centering
\includegraphics[width=9cm]{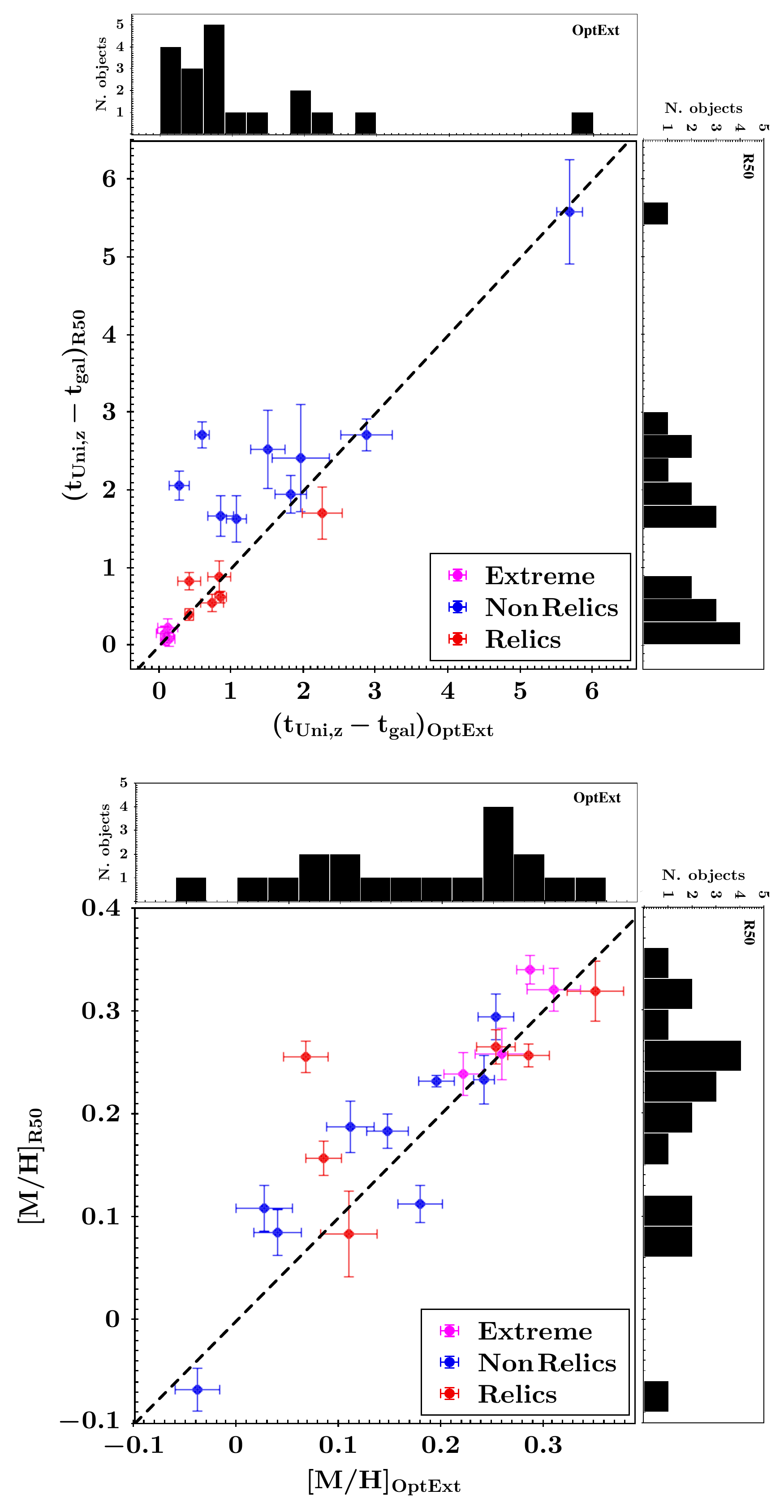}
\caption{Scatter plots  of the formation times (top) and metallicities (bottom) derived from the OptExt (x-axis) and the R50 (y-axis) for the \INSPIRE\, DR1 objects, which are colour coded according to their relic class (Sect.~\ref{sec:results_confirmation}). In all panels, we always plot the mean between the values obtained with REG=0 and with the MAX\_REGUL, and show the one-to-one line (dashed black). The bins in the histograms are chosen to be as similar as possible to the uncertainties on the relative quantities.  }
\label{fig:stel_pop_distribution}
\end{figure}

In Figure~\ref{fig:stel_pop_distribution} we show the scatter plots of the formation times and metallicities obtained with \ppxf\, from the OptExt (x-axis) and in the R50 (y-axis) spectra. We show instead the distribution of these quantities for the two extraction methods separately in the histograms on the side of and at the top of each panel.  We do not directly plot the ages in the top panel, because the galaxies are at different redshifts. 
Instead, we use the formation time defined as the difference between the age of the Universe at the redshift of the object and the mass-weighted age derived from full spectral fitting $(t_{\text{Uni,z}} - t_{\text{gal}})$. This quantity is a proxy for the formation age of the systems: the closer this number is to zero, the earlier the object assembled
its stellar mass in the history of the Universe. Here,  $t_{\text{gal}}$ indicates the mean value between the unregularised and the regularised fit, which are reported in the fifth and sixth columns of Table~\ref{tab:stelpop}. We stress that the R50 and OptExt estimates do not necessarily have to be on the one-to-one line because the two extractions cover different spatial regions of the galaxies and give different weights to the different pixels on the slit (i.e. the  OptExt extraction is weighted by flux). This is indeed what happens for non-relic galaxies (blue points), and we return to this point in Section~\ref{sec:results_confirmation}.  Briefly, depending on how much the possible `accreted' populations contribute to the light, and at which radius, we can have overall older or younger ages and higher or lower metallicites for one or the other extraction method. The difference are instead expected to be negligible for relics (red and magenta points in the figure), which are made almost entirely of in situ stars. 

\begin{table*}
\centering

\begin{tabular}{lcc|crrrrrrr|r}
\hline
\hline
\multicolumn{1}{c}{ID} &
\multicolumn{1}{c}{t$_{\text{Uni, z}}$} &
\multicolumn{1}{c|}{Ext.} &
\multicolumn{1}{c}{MAX} &
\multicolumn{1}{c}{t$_{\text{REG=0}}$} &
\multicolumn{1}{c}{t$_{\text{RMAX}}$} &
\multicolumn{1}{c}{$\delta$Age} &
\multicolumn{1}{c}{Z$_{\text{REG=0}}$} &
\multicolumn{1}{c}{Z$_{\text{RMAX}}$} &
\multicolumn{1}{c}{$\delta$Z} &
\multicolumn{1}{c}{$\sigma_{\star\text{, st.pop.}}$} &
\multicolumn{1}{|c}{[Mg/Fe]} \\
\multicolumn{1}{c}{KiDS} &
\multicolumn{1}{c}{(Gyrs)} &
\multicolumn{1}{c|}{Method} &
\multicolumn{1}{c}{REGUL} &
\multicolumn{1}{c}{(Gyr)} &
\multicolumn{1}{c}{(Gyr)} &
\multicolumn{1}{c}{(Gyr)} &
\multicolumn{1}{c}{ } &
\multicolumn{1}{c}{} &
\multicolumn{1}{c}{} &
\multicolumn{1}{c}{(km s$^{-1}$)} &
\multicolumn{1}{|c}{} \\
\hline
\hline
\multirow{2}{*}{J0211-3155} & \multirow{2}{*}{10.3} & OptExt & 31.0   & 10.5 & 10.0  & 0.2  & +0.22 & +0.23 & 0.02 & $236\pm4$  &  $0.2\pm0.1$ \\
                            &                       &R50     & 36.0   & 10.5 &  9.8  & 0.2  & +0.23 & +0.25 & 0.02 & $234\pm4$  &  $0.3\pm0.1$ \\
\multirow{2}{*}{J0224-3143} & \multirow{2}{*}{9.5}  & OptExt & 6.0    &  9.1 &  8.5  & 0.2  & +0.24 & +0.26 & 0.02 & $246\pm1$  &  $0.3\pm0.1$ \\
                            &                       &R50     & 5.5    &  9.1 &  8.8  & 0.2  & +0.25 & +0.28 & 0.02 & $250\pm2$  &  $0.3\pm0.1$ \\
\multirow{2}{*}{J0226-3158} & \multirow{2}{*}{10.9} & OptExt & 46.0    & 10.6 & 9.9  & 0.2  & +0.15 & +0.21 & 0.02 & $182\pm2$  &  $0.2\pm0.1$ \\
                            &                       &R50     & 3000.0  &  8.5 &  7.9  & 0.2  & +0.10 & +0.12 & 0.02 & $184\pm2$  &  $0.0\pm0.1$\\
\multirow{2}{*}{J0240-3141} & \multirow{2}{*}{10.5} & OptExt & 1092.5 &  8.8 &  8.2  & 0.2  & +0.26 & +0.22 & 0.01 & $219\pm2$  &  $0.0\pm0.1$ \\
                            &                       &R50     & 115.0  &  9.1 &  7.1  & 0.3  & +0.27 & +0.30 & 0.02 & $217\pm1$  &  $0.0\pm0.1$\\
\multirow{2}{*}{J0314-3215} & \multirow{2}{*}{10.4} & OptExt & 2831.0 &  8.7 &  6.4  & 0.3  & +0.13 & +0.17 & 0.02 & $179\pm3$  &  $0.1\pm0.1$ \\
                            &                       &R50     & 27.0   &  7.6 &  7.8  & 0.2  & +0.18 & +0.19 & 0.02 & $171\pm2$  &  $0.2\pm0.1$\\
\multirow{2}{*}{J0316-2953} & \multirow{2}{*}{9.7}  & OptExt & 50.0   &  7.7 &  7.2  & 0.4  & +0.11 & +0.11 & 0.03 & $169\pm3$  &  $0.3\pm0.1$ \\
                            &                       &R50     & 14.0   &  8.3 &  7.7  & 0.4  & +0.09 & +0.07 & 0.04 & $173\pm4$  &  $0.4\pm0.1$\\
\multirow{2}{*}{J0317-2957} & \multirow{2}{*}{10.6} & OptExt & 12.0   & 10.5 &  9.9  & 0.2  & +0.04 & +0.09 & 0.02 & $161\pm3$  &  $0.4\pm0.1$ \\
                            &                       &R50     & 11.5   & 10.2 &  9.3  & 0.2  & +0.23 & +0.28 & 0.01 & $168\pm3$  &  $0.3\pm0.1$\\
\multirow{2}{*}{J0321-3213} & \multirow{2}{*}{10.3} & OptExt & 39.0   & 10.4 &  9.6  & 0.3  & +0.01 & +0.04 & 0.03 & $190\pm3$  &  $0.4\pm0.1$ \\
                            &                       &R50     & 24.0   &  8.3 &  8.1  & 0.3  & +0.13 & +0.08 & 0.02 & $190\pm3$  &  $0.3\pm0.1$\\
\multirow{2}{*}{J0326-3303} & \multirow{2}{*}{10.3} & OptExt & 62.0   &  9.6 &  9.3  & 0.2  & -0.06 & -0.01 & 0.02 & $156\pm3$  &  $0.2\pm0.1$ \\
                            &                       &R50     & 81.0   &  8.7 &  8.6  & 0.3  & -0.09 & -0.05 & 0.02 & $162\pm3$  &  $0.2\pm0.1$\\
\multirow{2}{*}{J0838+0052} & \multirow{2}{*}{10.6} & OptExt & 10.0   & 10.1 &  9.4  & 0.2  & +0.09 & +0.09 & 0.02 & $168\pm3$  &  $0.4\pm0.1$ \\
                            &                       &R50     & 9.0    & 10.1 &  9.3  & 0.2  & +0.15 & +0.16 & 0.02 & $173\pm3$  &  $0.3\pm0.1$ \\
\multirow{2}{*}{J0842+0059} & \multirow{2}{*}{10.3} & OptExt & 27.0   & 10.5 &  9.9  & 0.2  & +0.24 & +0.28 & 0.03 & $297\pm8$  &  $0.4\pm0.1$ \\
                            &                       &R50     & 27.0   & 10.5 &  9.9  & 0.2  & +0.24 & +0.28 & 0.03 & $285\pm8$  &  $0.4\pm0.1$\\
\multirow{2}{*}{J0847+0112} & \multirow{2}{*}{11.5} & OptExt & 6.5    & 11.5 & 11.2  & 0.2  & +0.27 & +0.31 & 0.01 & $224\pm3$  &  $0.3\pm0.1$ \\
                            &                       &R50     & 26.5   & 11.4 & 11.1  & 0.2  & +0.32 & +0.36 & 0.01 & $235\pm2$  &  $0.2\pm0.1$\\
\multirow{2}{*}{J0857-0108} & \multirow{2}{*}{10.6} & OptExt & 13.5   &  9.8 &  9.2  & 0.2  & +0.18 & +0.21 & 0.02 & $160\pm3$  &  $0.2\pm0.1$ \\
                            &                       &R50     & 299.0  &  9.3 &  8.7  & 0.2  & +0.25 & +0.21 & 0.01 & $167\pm2$  &  $0.1\pm0.1$ \\
\multirow{2}{*}{J0918+0122} & \multirow{2}{*}{9.6}  & OptExt & 93.5   &  8.1 &  7.4  & 0.2  & +0.25 & +0.26 & 0.02 & $233\pm3$  &  $0.1\pm0.1$ \\
                            &                       &R50     & 34.5   &  8.0 &  7.3  & 0.2  & +0.26 & +0.32 & 0.02 & $242\pm3$  &  $0.1\pm0.1$\\
\multirow{2}{*}{J0920+0212} & \multirow{2}{*}{10.5} & OptExt & 49.0   &  9.7 &  9.5  & 0.2  & +0.29 & +0.29 & 0.02 & $225\pm4$  &  $0.2\pm0.1$ \\
                            &                       &R50     & 7.0    & 10.3 &  9.5  & 0.2  & +0.24 & +0.27 & 0.01 & $230\pm3$  &  $0.2\pm0.1$\\
\multirow{2}{*}{J2305-3436} & \multirow{2}{*}{10.3} & OptExt & 17.5   & 10.4 &  9.9  & 0.2  & +0.28 & +0.34 & 0.03 & $313\pm1$  &  $0.4\pm0.1$ \\
                            &                       &R50     & 19.0   & 10.5 & 10.0  & 0.2  & +0.29 & +0.35 & 0.02 & $309\pm1$  &  $0.3\pm0.1$ \\
\multirow{2}{*}{J2312-3438} & \multirow{2}{*}{9.7}  & OptExt & 18.0   &  8.6 &  7.8  & 0.3  & +0.03 & +0.05 & 0.02 & $210\pm3$  &  $0.3\pm0.1$ \\
                            &                        &R50    & 18.0   &  7.6 &  6.8  & 0.4  & +0.08 & +0.09 & 0.02 & $220\pm1$  &  $0.2\pm0.1$  \\
\multirow{2}{*}{J2327-3312} & \multirow{2}{*}{9.3}  & OptExt & 3000.0 &  4.6 &  2.7  & 0.5  & +0.12 & +0.11 & 0.02 & $214\pm2$  &  $0.1\pm0.1$ \\
                            &                       &R50     & 63.0   &  3.9 &  3.6  & 0.5  & +0.20 & +0.18 & 0.02 & $213\pm2$  &  $0.2\pm0.1$\\
\multirow{2}{*}{J2359-3320} & \multirow{2}{*}{10.4} & OptExt & 45.5   & 10.4 &  9.5  & 0.2  & +0.35 & +0.34 & 0.03 & $221\pm3$  &  $0.2\pm0.1$ \\
                            &                       &R50     & 45.0   & 10.4 &  9.6  & 0.2  & +0.31 & +0.32 & 0.03 & $219\pm4$  &  $0.2\pm0.1$\\
\hline
\hline
\end{tabular}
\caption{Stellar population parameters of the \INSPIRE\, DR1 galaxies, obtained from the spectra extracted with the two different methods. The second column reports the age of the Universe at the redshift of the systems. 
The last column lists the [Mg/Fe] values inferred from line indices, while mass-weighted ages and metallicites, provided in the central block of columns, are obtained from the \ppxf\, fit, in the unregularised and regularised cases, as described in the text. The uncertainties associated with these measurements include only the error budget associated with the observational noise in the assumption of independent Gaussian errors on each pixel. This is estimated repeating the fit multiple times, changing the configuration, the setup and the regularisation of the fit (always kept between 0 and MAX\_REGUL, also listed in the table). Finally, we also give the estimated stellar velocity dispersion obtained during the stellar population \ppxf\, fit.  The new values are fully consistent with the previous ones derived in Section~\ref{sec:kinematics}.}     \label{tab:stelpop}
\end{table*}

\subsection{ Relic confirmation}
\label{sec:results_confirmation}
Having inferred mass-weighted integrated ages, we can now identify the systems that formed the majority of their stellar mass during the first phase of the two-phase formation scenario, and  
thus confirm or reject the relic nature of the 19 objects in the \INSPIRE\, DR1. According to the two-phase formation scenario, relics are systems that were already almost fully  assembled at the end of the first phase (assumed here to be $z\sim2$, corresponding to 3 Gyr after the BB; see e.g.  \citealt{Zolotov15}). It therefore seems natural to define as relics those objects for which the great majority of the cumulative stellar mass fraction was already in place by that time. 

As already suggested by \citetalias{Ferre-Mateu+17} a `degree of relicness' can be defined to 
quantify how `extreme' the structural parameters, the mass of the black hole \citep{Ferre-Mateu+15}, and the star formation history (SFH) of each object are.
Now, for the first time, we have at our disposal a large number of confirmed UCMGs with which to test the existence of such a degree of relicness as a function of the SFH, especially when considering all the possible setups and uncertainties in the fit. This helps us to build an operative definition of `relic' which can also then be compared to other galaxy properties, such as size, mass, [Mg/Fe], and local environment. 

We compute, for each  object and for each extraction method, the percentage of  stellar mass assembled as a function of cosmic time, for the two extreme \ppxf\, configurations (i.e. unregularised and REGUL = MAX$\_$REGUL) and also considering the uncertainties on  the [$\alpha$/Fe] of the SSP models we use for the fitting\footnote{We remind the readers that we run \ppxf\, using models with [$\alpha$/Fe] equal to the [Mg/Fe] measured from indices, but we also run the fit considering models with values equal to [Mg/Fe]$\pm0.1$.}. 
We then take, for each spectrum, the minimum value (among the various ones inferred with the different assumptions in the \ppxf\, fit) for the stellar mass assembled 3 Gyr after the BB (M$_{\star, t_{\text{BB=3 Gyr}}}$).  
These values for the 19 objects presented in this \INSPIRE\, DR1 are listed in Table~\ref{tab:degree_of_relicness}. In addition to M$_{\star, t_{\text{BB=3 Gyr}}}$, in the first block of columns of the table, we report the time of final assembly ($t_{\text{fin}}$), that is, the cosmic time from the BB at which the given object has assembled more than 95\% of its stellar mass, and the time at which 75\% of M$_{\star}$ was already in place. 
For all these quantities, we obtain the mean between the OptExt and R50 extraction values, while for completeness we report the single values for each extraction in the second block. For seven systems (including one relic) the values inferred from the two extractions differ significantly.  

Table~\ref{tab:degree_of_relicness} allows us to define a 
`degree of relicness' in the SFH, whereby the first system, J0847+0112 had already fully assembled at $z=2$ (i.e. 100\% of the stellar mass was already in place), while J2327-3312 had barely begun making stars at that time.  We then set up two thresholds (highlighted as horizontal lines in the Table) on the <M$_{\star,t_{\text{BB}}=3 \text{Gyr}}$> and the time of assembly, in order to split the 19 systems into three main groups: 
\begin{itemize}
    \item \textit{Extreme relics:} objects that have fully assembled their stellar mass by $z=2$ (M$_{\star,t_{\text{BB}}=3 \text{Gyr}} >0.99$ and  $t_{\text{fin}}$<3 Gyr).    
    \item \textit{Relics:} objects that have assembled 75\% or more of their stellar mass by $z=2$ (M$_{\star,t_{\text{BB}}=3\text{Gyr}} >0.75$, or equivalently $t_{75\%}<3$ Gyr) in both the OptExt and R50 extractions.  
    \item \textit{Non-relics:} everything else, with less than 75\% of the stellar mass already in place 3 Gyr after the BB in at least one extraction. These objects are characterised by a more extended SFH, which finished at later cosmic times, and,  despite their very small sizes, show the presence of an `accreted' population (including stars brought in from merger events, but also these formed later on during subsequent star formation episodes). This can be deduced from the bottom panels of the figures, showing the 2D density plot in a $\log(age)$--metallicity space.
\end{itemize}

Following this simple scheme, among the 19 systems analysed here and released as part of \INSPIRE\, DR1, we find a total of 10 systems (including 2 already presented in the Pilot) that had already assembled more than 85\% (although the threshold was fixed to 75\%) of their stellar mass by the end of the compaction phase at $z\sim2$. 
Among these, 4 are `extreme relics' as they were completely formed after the first phase of the formation and did not experience any interaction and/or accretion nor any other star formation after the first, very quick mass assembly episode. 

\begin{table*}
\begin{tabular}{lccc|ccccccl}
\hline
\hline
\multicolumn{1}{c}{ID KiDS} & 
\multicolumn{1}{c}{<M$_{\star,t_{\text{BB}}=3}$>} & 
\multicolumn{1}{c}{<t$_{\text{fin}}$> } &
\multicolumn{1}{c|}{<t$_{75\%}$> } &
\multicolumn{1}{c}{M$_{\star,t_{\text{BB}}=3}$} &
\multicolumn{1}{c}{M$_{\star,t_{\text{BB}}=3}$} & 
\multicolumn{1}{c}{t$_{\text{fin}}$ (Gyr)} &
\multicolumn{1}{c}{t$_{\text{fin}}$ (Gyr)} &
\multicolumn{1}{c}{t$_{75\%}$ (Gyr)} & 
\multicolumn{1}{c}{t$_{75\%}$ (Gyr)} &  \multicolumn{1}{c}{Class}  \\
&
\multicolumn{1}{c}{ } & 
\multicolumn{1}{c}{(Gyr) } &
\multicolumn{1}{c|}{(Gyr) } &
\multicolumn{1}{c}{OptExt} &
\multicolumn{1}{c}{R50}    & 
\multicolumn{1}{c}{OptExt} &
\multicolumn{1}{c}{R50}    & 
\multicolumn{1}{c}{OptExt} &
\multicolumn{1}{c}{R50} & 
\multicolumn{1}{c}{ } \\
\hline
  J0847+0112 & 1.000   & 1.75 & 1.00  & 1.0  & 1.0  &  1.5 & 2.0 & 1.0 & 1.00 & extreme\\
  J0842+0059 & 1.000   & 2.00  & 1.50  & 1.0  & 1.0  &  2.0 & 2.0 & 1.5 & 1.5 & extreme\\   J2305-3436 & 0.995 & 2.00  & 1.50  & 0.99 & 1.0  &  2.0 & 2.0 & 1.5 & 1.5 & extreme\\
  J0211-3155 & 0.995 & 2.00  & 1.50  & 1.00  & 0.99 &  2.0 & 2.0 & 1.5 & 1.5 & extreme \\
  \hline
  J0224-3143 & 0.975 & 2.50  & 1.25 & 0.97 & 0.98 &  3.0 & 2.0 & 1.5 & 1.0 & relic \\
  J2359-3320 & 0.975 & 2.75 & 2.00  & 0.97 & 0.98 &  3.0 & 2.5 & 2.0 & 2.0 & relic \\
  J0920+0212 & 0.960  & 3.00  & 1.50  & 0.97 & 0.95 &  3.0 & 3.0 & 1.5 & 1.5 & relic \\  
  J0317-2957 & 0.945 & 3.25 & 1.50  & 0.99 & 0.9  &  2.5 & 4.0 & 1.5 & 1.5 & relic \\
  J0838+0052 & 0.955 & 4.00  & 2.25 & 0.99 & 0.92 &  3.0 & 5.0 & 2.0 & 2.5 & relic \\
  J0316-2953$*$ & 0.870  & 6.50  & 2.00  & 0.84 & 0.9  &  9.0 & 4.0 & 2.0 & 2.0 & relic \\
  \hline
  J0857-0108$*$ & 0.795 & 7.00  & 4.50  & 0.87 & 0.72 &  6.0 & 8.0 & 2.0 & 7.0 & non-relic \\
  J0918+0122 & 0.770  & 8.50  & 3.25 & 0.72 & 0.82 &  8.5 & 8.5 & 3.5 & 3.0 & non-relic \\
  J0326-3303$*$ & 0.735 & 4.75 & 2.75 & 0.94 & 0.53 &  4.0 & 5.5 & 1.5 & 4.0 & non-relic \\
  J2312-3438 & 0.690  & 8.25 & 3.25 & 0.79 & 0.59 &  8.0 & 8.5 & 2.5 & 4.0 & non-relic \\
  J0226-3158$*$ & 0.685 & 6.25 & 3.75 & 0.79 & 0.58 &  5.0 & 7.5 & 2.0 & 5.5 & non-relic \\
  J0321-3213$*$ & 0.645 & 5.75 & 2.75 & 0.95 & 0.34 &  2.5 & 9.0 & 1.5 & 4.0 & non-relic \\
  J0240-3141$*$ & 0.645 & 7.50  & 4.00  & 0.74 & 0.55 &  5.5 & 9.5 & 3.0 & 5.0  & non-relic\\
  J0314-3215 & 0.620  & 9.50  & 4.50  & 0.52 & 0.72 &  9.5 & 9.5 & 5.5 & 3.5 & non-relic \\
  J2327-3312$*$ & 0.275 & 9.00  & 8.50  & 0.14 & 0.41 &  9.0 & 9.0 & 8.5 & 8.5  & non-relic\\
\hline
\hline
\end{tabular}
\begin{flushright}
\footnotesize{$^*$ For these objects, the percentages of mass and the times of assembly estimated from the R50 and from the OptExt differ substantially. }
\end{flushright}
\caption{Lower limit for the percentage of stellar mass already assembled 3 Gyr after the BB (i.e. at the end of the first phase of the formation), cosmic time of final assembly, and time at which 75\% of the stellar mass was already in place. The quantities are computed as the mean of the quantities calculated for the OptExt and R50 cases separately, which are listed in the second block of columns. The \INSPIRE\, systems classified in the three groups identified by the last column (and separated by horizontal lines) are listed in descending order of degree of relicness, from the most extreme, to the galaxy with the most extended and recent SFH. }
\label{tab:degree_of_relicness}
\end{table*}

Furthermore, we note that for seven systems, of which only one is a relic (but the least `extreme' one, J0316-2953),  the time of completed assembly and/or the percentage of stellar mass assembled by the end of the first phase inferred from R50 and that from OptExt strongly differ. This likely happens because, by extracting the 1D spectra with a different method, we are integrating the light coming from different regions (also giving a different weight to the different regions). The differences in mass-weighted ages and metallicity would therefore reflect the presence of spatial gradients\footnote{Although we caution the readers that we cannot obtain spatially resolved information, nor investigate the stellar populations of these objects at very larger radii.}. 
For J0316-2953, the disagreement originates from a very late $t_{\text{fin}}$ in the OptExt case. However, we note that for that spectrum, 90\% of the stellar mass is already in place by $z=2$ and only the remaining 10\% is assembled over a very long cosmic time. We therefore classify this object as a relic, because the great majority of its mass was formed during the first phase of formation via a quick burst, given the very high [Mg/Fe].  For all the confirmed relics, the stellar population parameters, as well as the time of final assembly and the fraction of mass assembled during the first phase, are very similar between the two  extractions and often even 
consistent within the errors. This is another argument favouring the uniform stellar population in these objects.

\begin{figure}[h]
    \centering
    \includegraphics[width=\columnwidth]{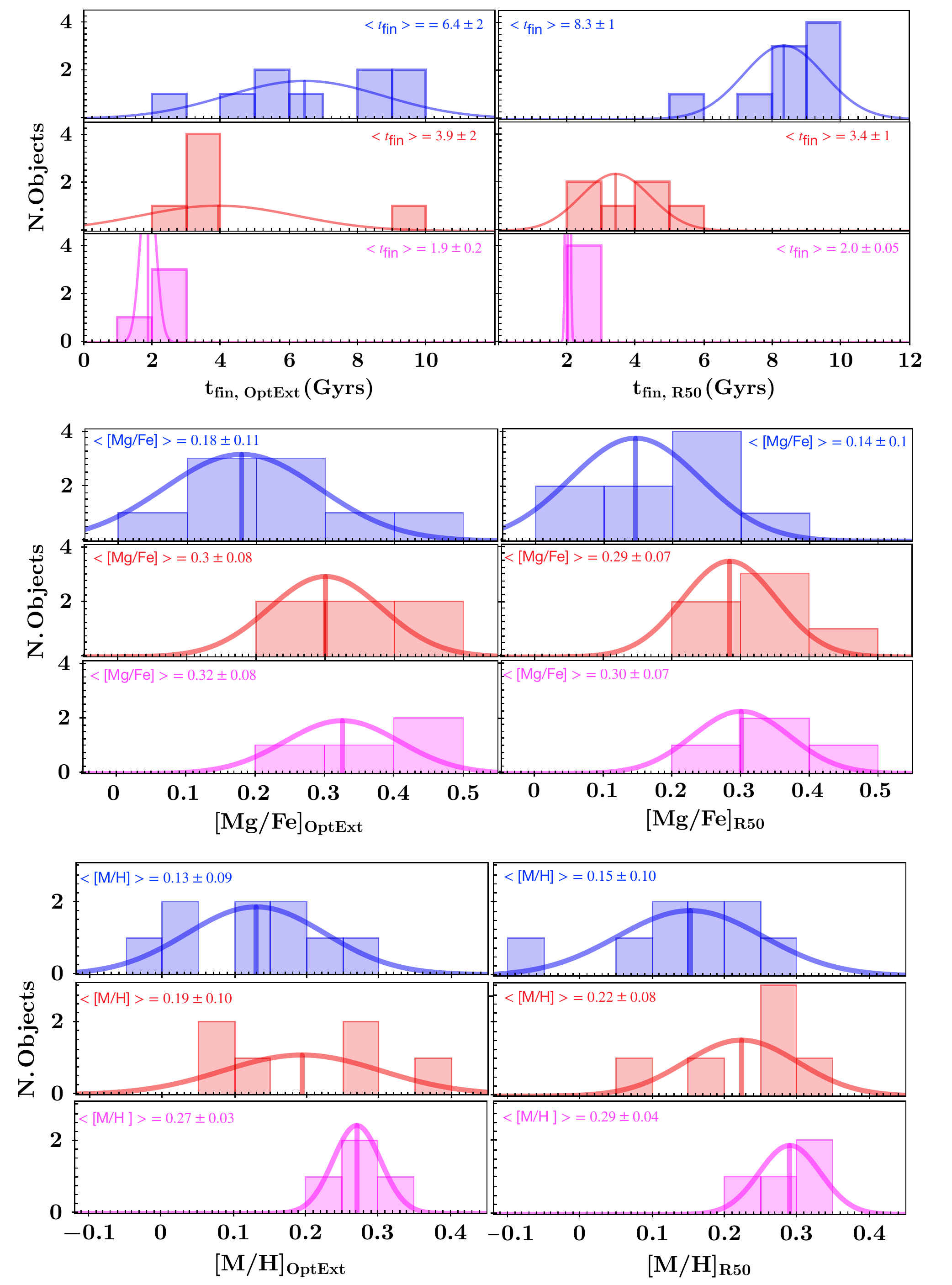}
    \caption{Distribution of time of final assembly (top), [Mg/Fe] (middle), and metallicities [M/H] (bottom) for the three classes, for the OptExt spectra (left) and the R50 ones (right). The average values of the Gaussian fits are also shown in each panel.}
    \label{fig:histo_mgFe_Z}
\end{figure}

Figure~\ref{fig:histo_mgFe_Z} shows the distribution of the time of final assembly, [Mg/Fe], and metallicities ([M/H]) for the three classes, namely non-relics (blue, top), relics (red middle), and extreme relics (magenta, bottom), obtained from both the OptExt (left) and the R50 (right) spectra. Gaussian fits to the histograms are overplotted on each panel and the mean values with standard deviation are given too. 
As expected, relics complete their assembly much earlier in cosmic time than non-relics ($t_{\text{fin, relics}}<<t_{\text{fin, non-relics}}$). Their star formation quenched in a shorter time, as demonstrated by the overall larger [Mg/Fe]. 
Finally, remarkably, there are no extreme relics with solar metallicity, although we note the poor statistics (four objects in total). 

\begin{figure*}
\centering
\includegraphics[width=18cm]{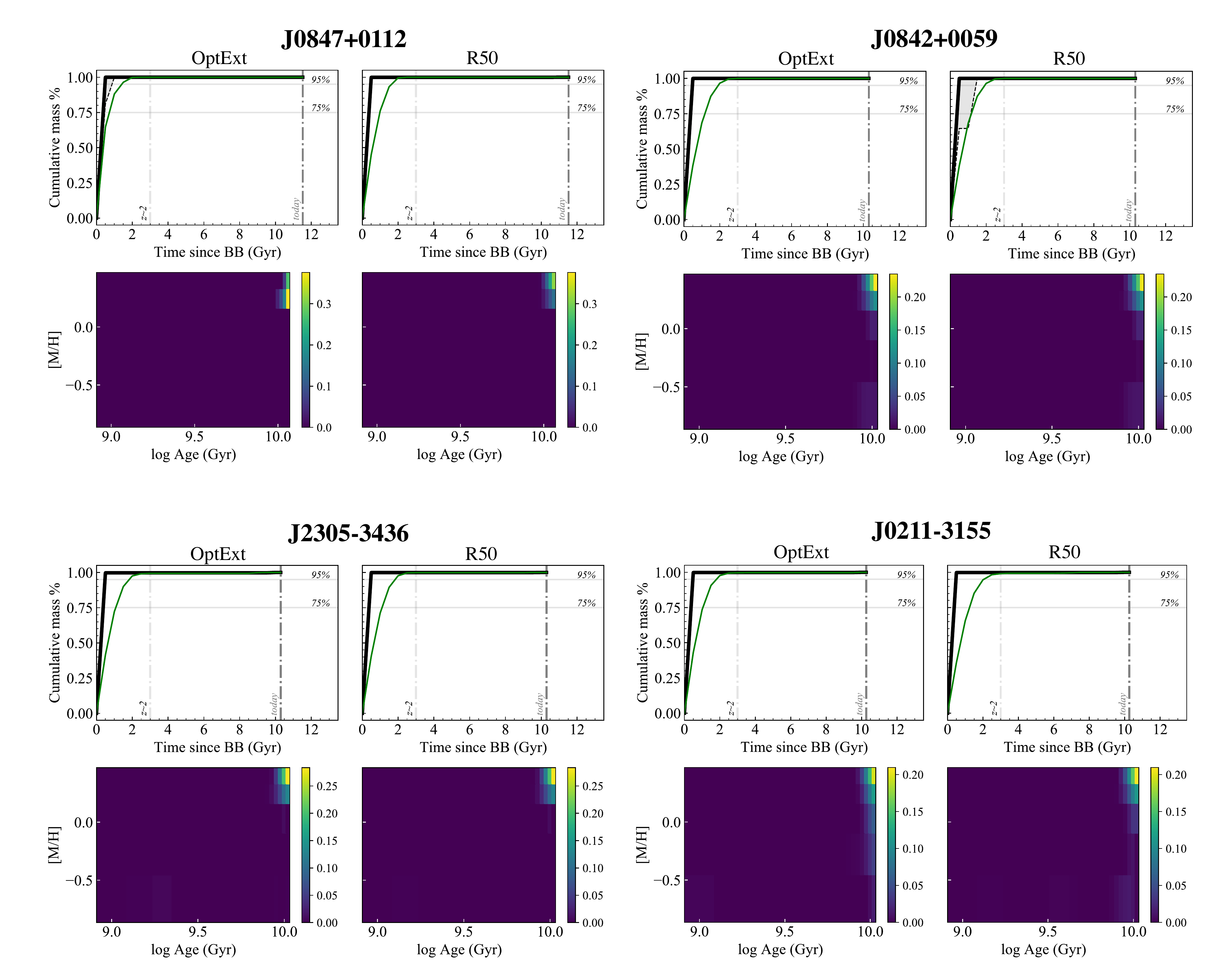}
\caption{Star formation history of the four extreme relics ($M_{\star}\ge$99\% at $t_{\text{BB=3 Gyr}}$). The systems are ordered in descending order of degree of relicness.  \textit{Upper plots of each panel:} cumulative stellar mass fraction formed in time, starting from the BB and up to the age of the Universe at the redshift of each system, which is plotted as a vertical dashed line. The other vertical line shows instead the 2 Gyr threshold used in the text to classify an object as an extreme relic. Black lines show the result obtained from the unregularised \ppxf\, fit with the best value of [$\alpha$/Fe] estimated from line indices. The shadow regions show the uncertainties on such estimates ($\pm0.1$). Finally, the green lines show the results of the fit performed with the regularisation set to the MAX\_REGUL. 
\textit{Bottom plots of each panel}: 2D density plot of the weights attributed to the SSP models by \ppxf\, in the $\log$ (age)--[M/H] space. The colour bar shows the weight fractions of the SSP models with the corresponding age and metallicity  used in the fit.}
\label{fig:stel_mass_extreme_relics}
\end{figure*}

\begin{figure*}
\centering
\includegraphics[width=18cm]{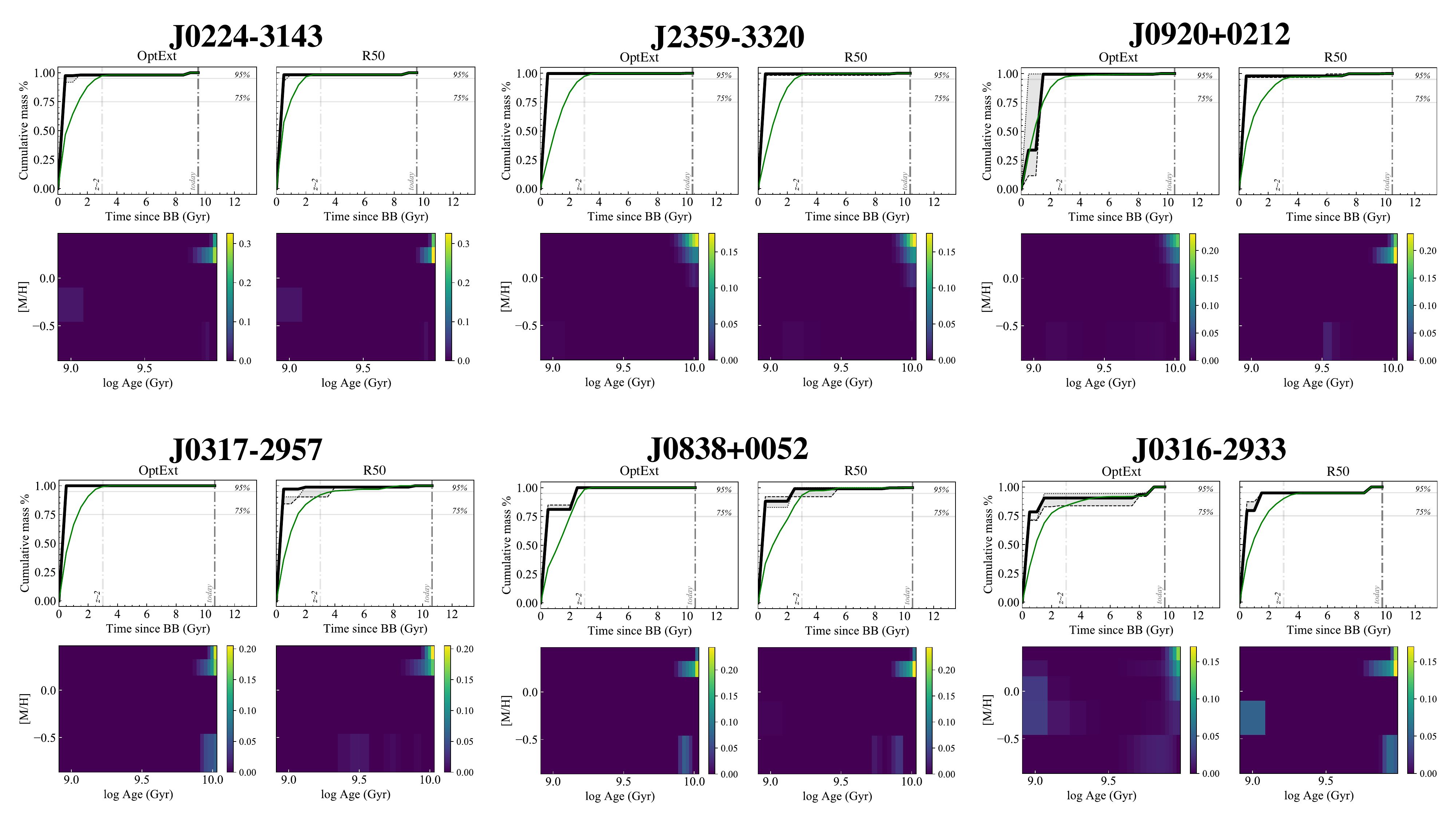}
\caption{Same as in Figure~\ref{fig:stel_mass_extreme_relics} but for relics (95\% >$M_{\star}\ge$75\% at $t_{\text{BB=3 Gyr}}$ and $t_{\text{75\%}}<3$ Gyr). The systems are plotted in order of decreasing  degree of relicness from top-left to bottom-right. Some indication for the presence of a second stellar population with younger ages and/or lower metallicity is visible in the density plots, especially in the bottom row, although the mass is still dominated by the pristine stars. }
\label{fig:stel_mass_relics}
\end{figure*}

\begin{figure*}
\centering
\includegraphics[width=18cm]{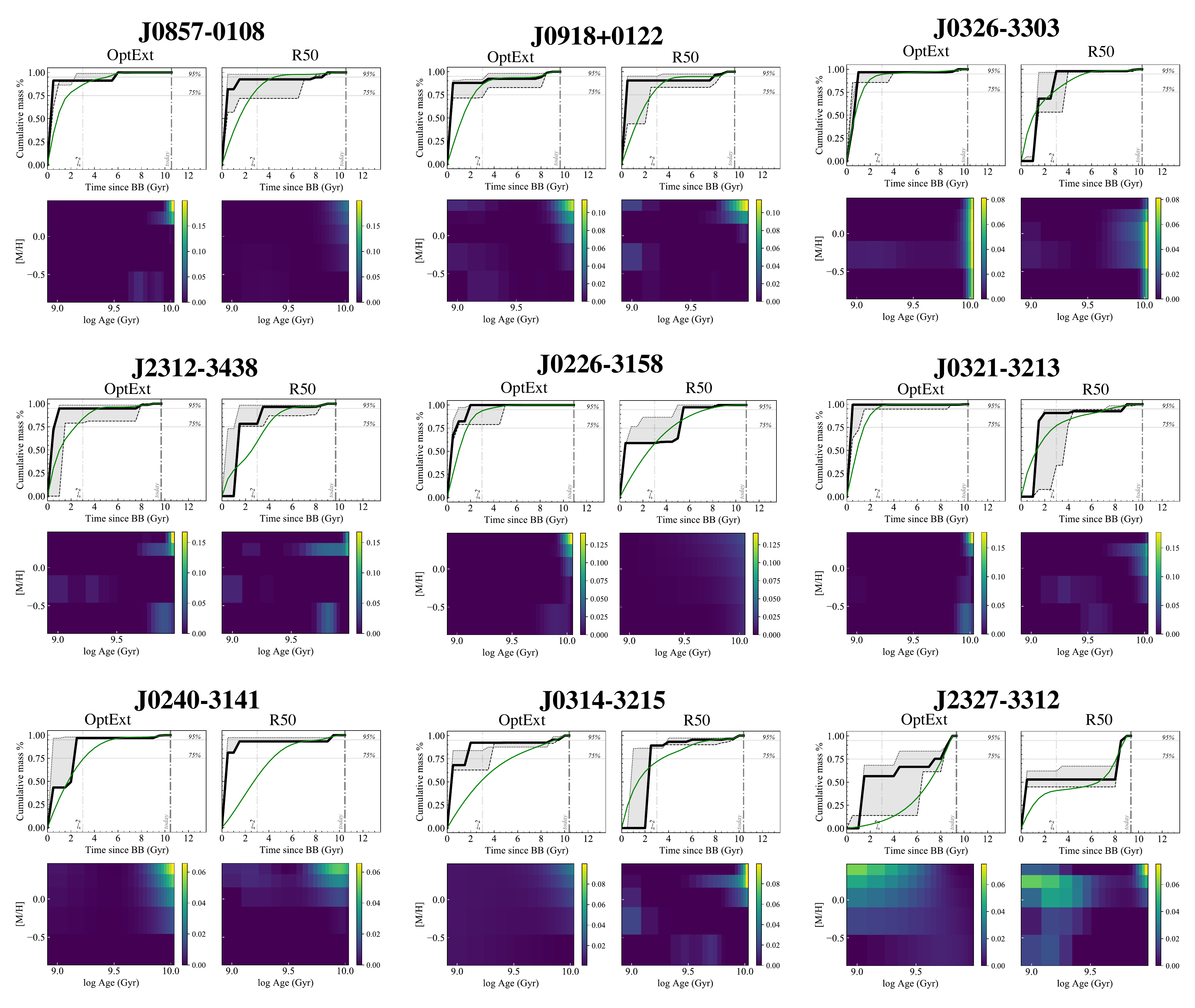}
\caption{Same as in Figures~\ref{fig:stel_mass_extreme_relics} and~\ref{fig:stel_mass_relics} but for the systems that are not relics.  
Clear indications for the presence of a second stellar population with younger ages and/or lower metallicity is visible in all the density plots and in the cumulative SFHs. }
\label{fig:stel_mass_not_relics}
\end{figure*}

A visualisation of the cumulative stellar mass as a function of cosmic time (from the BB to the age of the Universe at the redshift of each object) is provided in Figures~\ref{fig:stel_mass_extreme_relics},~\ref{fig:stel_mass_relics}, and~\ref{fig:stel_mass_not_relics}, for the three groups defined above. We caution the readers that the SFHs derived with \ppxf\, are non-parametric and are represented by a series of bursts (idealised as SSPs of different ages and metallicities). This approach and the large number of SSPs adopted ($\sim100$) provide the maximum degree of flexibility. This is a clear advantage over assuming {\it a priori} a simple parametric SFH, especially if we want to be able to trace a two-phase formation scenario, separating the stars formed in situ and those formed later on from accreted material (if present, i.e. for non-relics). Hence, in the case of relics, where almost the totality of the stars formed during the first phase, the star formation history would effectively be a single burst, while for non-relics, \ppxf\, uses multiple SSPs with very different ages (i.e. multiple bursts occurring at different cosmic times).

\subsection{Preliminary lower limit to the number density of relics from \INSPIRE\,}
\label{sec:results_numbdens}
Based on the number of confirmed relics ($N_{conf}= 10$),  we estimate a conservative lower limit on the number density of relics at $0.17<z_{\text{rel}}<0.39$. 
To obtain this, we follow the approach undertaken in \citetalias{Tortora+18_UCMGs}: we first consider the co-moving volume corresponding to the total sky area that was used to search for the UCMGs from which the \INSPIRE\ sample is drawn (KiDS DR3, 333 deg$^2$, \citealt{deJong+17_KiDS_DR3}). 
Specifically, $f_{\text{area}}=A_{\text{sky}}/A_{\text{KiDS}}$, where $A_{\text{sky}} = 41253$ deg$^{2}$ is the full sky area and $A_{\text{KiDS}}=333$ deg$^2$. 
We then compute the cosmic volume contained within the redshift window in which the confirmed relics are found:  $V_{\text{KiDS}} = f_{\text{area}}(V_{\text{z=0.39}}-V_{\text{z=0.17}})$.
At this point, the number density of relics estimated from \INSPIRE\, DR1 will simply be equal to 
\begin{equation}
\rho_{DR1} = \dfrac{N_{\text{conf}}}{V_{\text{KiDS,z}_{\text{rel}}}} =  9.1 \times 10^{-8} \text{Mpc}^{-3}  
,\end{equation}
which is about $0.5$ orders of magnitude lower than the value obtained in \citetalias{Ferre-Mateu+17} ($\sim6 \times 10^{-7} \text{Mpc}^{-3}$). This is not surprising, as we have not analysed the full list of candidates yet (but we have used the whole volume in the equation). Indeed, extrapolating the number density computation to what we will potentially find once \INSPIRE\ is completed (over the same sky area), assuming a successful rate for the relic confirmation of $\sim53$\% (i.e.  10/19), 
we obtain $\rho \sim 2 \times 10^{-7} Mpc^{-3}$, which is closer to the value reported by \citetalias{Ferre-Mateu+17} and to the predictions from simulations. We stress that this number has to be considered as a lower limit on the relic number density, as the completeness and the selection function of the UCMGs have not been taken into account.  

\subsection{Do relics have larger velocity dispersions than non-relics at equal stellar mass?}
\label{sec:results_sigma-M}
An interesting and somewhat unexpected result is that there seems to be no strong correlation between the degree of relicness ---defined above according to stellar mass assembled in the first phase and the cosmic time of final assembly---  
and  global galaxy parameters such as stellar mass, size, and colours. However, at the moment, the statistics is still too poor to draw secure conclusions. We will therefore come back to this point once the final \INSPIRE\, sample has been fully analysed. 

\begin{figure}
\includegraphics[width=\columnwidth]{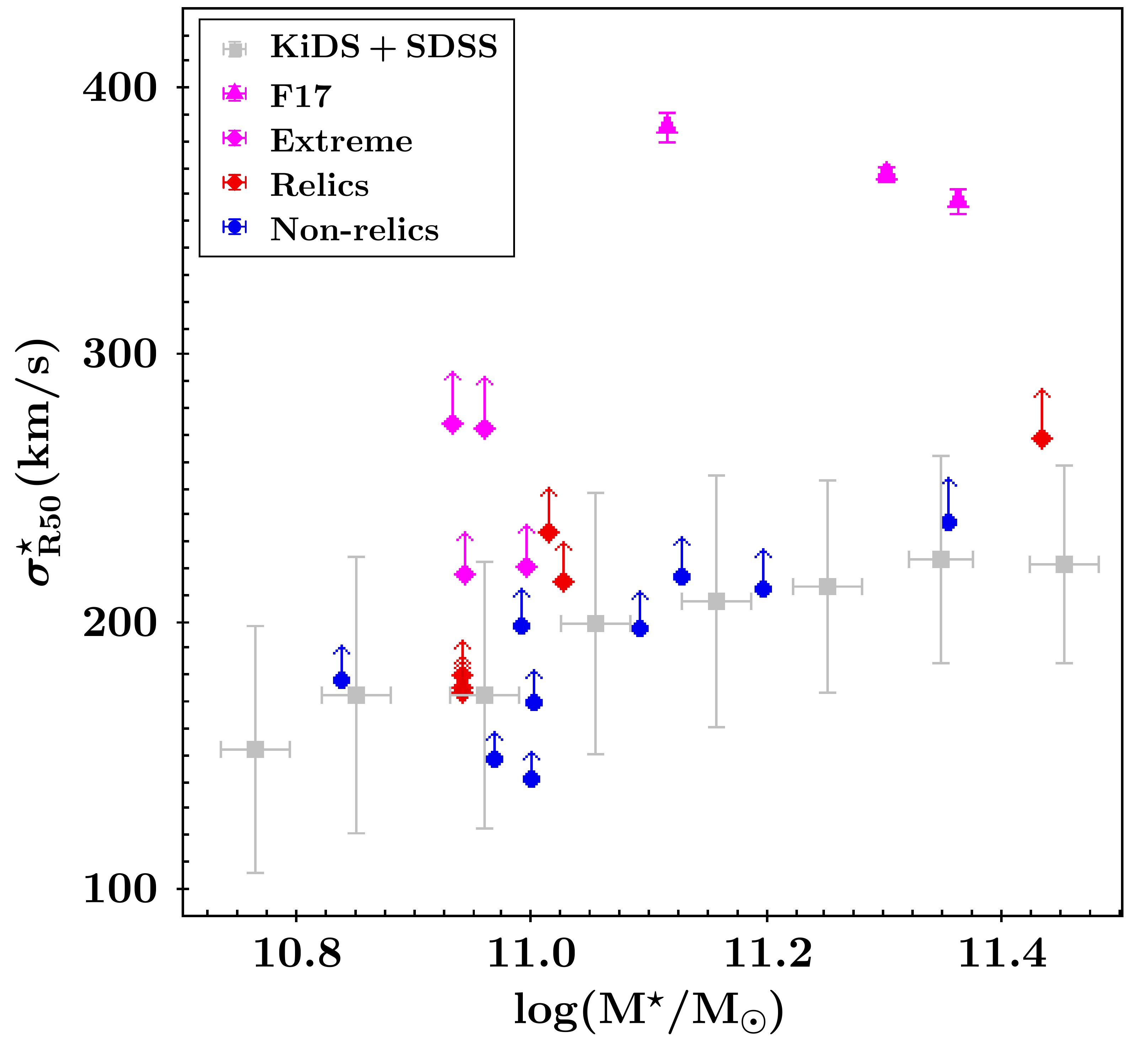}
\caption{Total stellar mass vs. stellar velocity dispersion for \INSPIRE\, relics (magenta and red diamonds for extreme and normal, respectively), and non-relics (blue circles). Grey squares are values obtained for normal-sized KiDS galaxies for which SDSS DR16 spectroscopy is available. Magenta  triangles are the three confirmed (extreme) relics in the local Universe (\citetalias{Ferre-Mateu+17}). For the \INSPIRE\, objects, the R50 velocity dispersion values have to be considered lower limits, because of the effect of the seeing (see Appendix~\ref{app:seeing_correction}). The arrow shows the averaged expected variation. }
\label{fig:mass-vdisp}
\end{figure}

The only notable quantitative difference between (extreme) relics and non-relics is visible in the stellar mass--stellar velocity dispersion 
parameter space, where we observe 
that at fixed stellar mass, relics have on average higher measured values of $\sigma_{\star}$ 
than non-relics. This is certainly true for extreme relics, including also the one in the local Universe \citepalias{Ferre-Mateu+17}. For normal relics,  the situation is not clear and a larger number of objects is necessary in order to draw conclusions. We visualise this in Figure~\ref{fig:mass-vdisp}, where we colour-code the objects according to their relic class, using the same colour code as in Figure~\ref{fig:histo_mgFe_Z}. The \citetalias{Ferre-Mateu+17} local relics are also included in the figure. These latter are extreme relics in terms of their SFH, and are therefore coloured in magenta, although with a different symbol (triangles).  
In the same figure, we overplot data on normal-sized KiDS DR3 galaxies, which have been selected to cover a similar range in redshift,$g-i$ colour, and stellar mass than the UCMGs. For these objects, the velocity dispersion has been computed by the Sloan Digital Sky Survey (SDSS) DR16  \citep{Ahumada20_sdssdr16}, which we then corrected to the effective radius using the formula in \citet{Cappellari+06}. Stellar masses were computed from KiDS multi-band photometry using the same code and setup as for the UCMGs (\citetalias{Scognamiglio20}). We bin the galaxies in stellar mass bins  ($\Delta\log(M^{\star}/M_{\odot})=0.1$ dex) and plot for each of them the median values with standard deviation as error bars. For the \INSPIRE\, objects, we plot the velocity dispersion values measured from the R50 spectra, which are underestimated because of the seeing effect, and thus are interpreted as a lower limit. The arrows indicate the average increase on the measured  $\sigma^{\star}_{R50}$ that would be seen after correcting for seeing broadening
(see Appendix~\ref{app:seeing_correction} for more details).

Although with limited statistics, we observe that the extreme relics (magenta symbols in Fig.~\ref{fig:mass-vdisp}) are systematically offset with respect to non-relics (blue circles) and normal-sized galaxies (grey squares) with the same stellar mass. 
The vertical offset is even larger for two of the four extreme relics, showing the highest $\sigma_{\star}$ ($\sim 280$ km/s) in the sample, and for the \citetalias{Ferre-Mateu+17} local relics.   
Using the virial theorem and thus assuming that  $\sigma_{\star}$ is approximately proportional to the dynamical mass of the galaxy within 1 \Reff\, \citep{SPIDER-VI}, the overall larger $\sigma_{\star}$ in relics would then indicate a higher dynamical mass compared to non-relic galaxies of similar stellar mass. 
This can have two possible physical explanations. 
One possibility is that relics could have a larger number of stars with very low masses, that is, a dwarf-rich IMF, which contribute very little to the light but substantially to the dynamical mass.  Another possibility is that relics could have a larger fraction of dark matter in the central regions or possibly dark matter halos with a higher central mass density. This is in line with the results reported in \citet{Gerhard01} 
who found that the dark  
halos in elliptical galaxies are much denser ($\sim 25$ times) than the halos of spiral galaxies of similar baryonic mass, suggesting a relation between the halo core density and the density of the Universe at the time of collapse (see also \citealt{Napolitano10,Thomas11b}), and therefore between the halo core density and the redshift of the collapse. As relics formed very early in cosmic time and evolved undisturbed and passively until today, it is likely that the physical properties of their dark matter halos have not been modified by later mass accretion or mergers. 

We note that this preliminary result has been inferred on a small number of objects. Our plan is to investigate these two scenarios further, also obtaining an  estimate of the IMF  slope from stellar population modelling,   and computing the fraction of internal dark matter from dynamical modelling upon availability of a larger sample of bona fide relic galaxies once the \INSPIRE\, LP is completed.

\section{Summary and Conclusions}
\label{sec:conclusions}
This paper presents the \INSPIRE\, DR1, which is also released as ESO Phase 3 collection\footnote{The 1D spectra extracted using the two different methods described here are available from the ESO Archive Science (\url{https://archive.eso.org/scienceportal/home?data_collection=INSPIRE}).}. We have reduced and analysed the X-Shooter spectra of the first 19 UCMGs selected from the KiDS Survey (\citetalias{Tortora+18_UCMGs, Scognamiglio20}) as good relic candidates. 

In particular we: 
\begin{itemize}
    \item[i)] extracted the 1D integrated spectra with two different methods, one encapsulating roughly 50\% of the light and one summing up all the light of the objects, but weighting the contribution by the flux at each pixel; 
    \item[ii)] obtained the stellar velocity dispersion from the UVB and from the VIS spectrum of each system, for each extraction method;
    \item[iii)] inferred light-weighted [Mg/Fe] from line-index strengths, and mass-weighted mean age and [M/H], performing full spectral fitting of the spectra with the \ppxf\, code; 
    \item[iii)] defined a `degree of relicness' computing the lower limit of the percentage of stellar mass already assembled at $z=2$, which corresponds to a Universe with an age of 3 Gyr (M$_{\star, t_{\text{BB=3 Gyr}}}$) and the time at which the galaxies had assembled 95\% and 75\% of their stellar masses;   
    \item[iv)] split the systems into three main groups, according to their M$_{\star, t_{\text{BB=3 Gyr}}}$, t$_{\text{fin}}$, and $t_{75\%}<3$ Gyr. 
    \item[v)] confirmed 10 (including the 2 from the pilot) of the 19 galaxies as relics (M$_{\star, t_{\text{BB = 3 Gyr}}}$>75\%), proving that they formed the majority of their mass ($>75$\%) with a short, high-z star formation episode (at $t_{BB}<3$ Gyr). All relics are characterised by a super-solar metallicity and high [Mg/Fe], similarly to the three confirmed local relics of \citetalias{Ferre-Mateu+17}, which provides further confirmation of the fact that their SFH extends over a very short timescale. 
    \item[vi)] identified, among these confirmed relics, four (including one from the pilot paper) as  `extreme-relics' with M$_{\star,t_{\text{BB=3 Gyr}}}>99$\% already 2 Gyr after the BB;
    \item[vii)] presented preliminary evidence for larger velocity dispersion values (and hence dynamical masses, assuming the virial theorem) in relics with respect to non-relics with similar stellar masses; 
    \item[viii)]  computed a preliminary lower limit for the number density of relics based on the objects confirmed in DR1 and extrapolating this number to the final \INSPIRE\, sample. Assuming a successful rate of $\sim50\%$, the lower limit on the number density of relics in the redshift window $0.17<z<0.39$  that we expect to measure from the complete sample is $\rho \sim 2 \times 10^{-7} Mpc^{-3}$, which is slightly lower than but broadly consistent with expectations from hydrodynamical simulations. 
\end{itemize}  

\section{Future data releases and scientific goals}
\label{sec:future}
In the second data release, foreseen in approximately six months, we will release the NIR spectra of these 19 systems. 
Extending the wavelength range towards the red, we will be able to constrain the low-mass end slope of the IMF. 
Given the more stringent requirements on S/N, we will stack the spectra of the objects confirmed as relics and those of the objects classified as non-relics and compare these two. According to \citetalias{Ferre-Mateu+17}, we should expect a measurable difference between the IMF slope of relics (expected to be bottom-heavy) and of non-relics that formed their stars with a more extended SFH and thus should have a more universal IMF. 

The third data release, which will be made available upon completion of all the spectroscopic observations, will be made of 52 objects, of which half will likely be confirmed as relics based on the current results (10/19). 
With more objects at our disposal, we will further investigate possible correlations between the degree of relicness and other galaxy properties (size, mass, colours). Thanks to the AMICO \citep{Maturi19} catalogue of galaxy cluster candidates from KiDS, we will also investigate whether a correlation between the degree of relicness and the local environment exists.  Finally, having a statistically large sample of relics, we will compare them with normal-sized ETGs with comparable stellar mass, redshifts, structural (S\'ersic index, $n$, axis ratio, $q$) and photometric (magnitudes and colours) parameters, and with optical spectra available.
In particular, for the normal-sized ETGs we will derive mass-weighted ages and metallicities and thus SFH using the same code (\ppxf) and under similar assumptions and setups (spectral resolution, multiplicative polynomial, regularisation). 
This will allow us to assess whether relics are simply the ultra-compact tail of the distribution of red and dead ETGs or are special and very rare systems with a completely different evolutionary path. 

\begin{acknowledgements}
CS is supported by an `Hintze Fellow' at the Oxford Centre for Astrophysical Surveys, which is funded through generous support from the Hintze Family Charitable  Foundation.  
CS, CT, FLB, AG, and SZ  acknowledge funding from the INAF PRIN-INAF 2020 program 1.05.01.85.11. AFM has received financial support through the Postdoctoral Junior Leader Fellowship Programme from `La Caixa' Banking Foundation (LCF/BQ/LI18/11630007). GD acknowledges support from CONICYT project Basal AFB-170002. 
DS is a member of the International Max Planck Research School (IMPRS) for Astronomy and Astrophysics
at the Universities of Bonn and Cologne.
\end{acknowledgements}

\bibliographystyle{aa} 
\bibliography{biblio_INSPIRE} 

\appendix

\section{Estimating the effects of the seeing on the velocity dispersion measurements}\label{app:seeing_correction}

Given the extremely small sizes of the UCMGs, the \INSPIRE\, 2D spectra are 
fully seeing dominated. The seeing is generally $\ge2$ times larger than the R50 defined in Section~\ref{subsec:1d-extr}, and 4-5 times larger than the effective radius of the \INSPIRE\, objects.
This means that, although by definition the 1D R50 spectra encapsulate 50\% of the total light, 
this is coming also from a region which is effectively larger than both $R_{e}$ and R50. 
Assuming that the velocity dispersion profiles decrease with increasing distance 
from the galaxies centre, we are measuring a lower integrated $\sigma_{\star}$ because of the seeing. 
As a consequence, the velocity dispersion values we computed from the R50 spectra 
in Section~\ref{sec:kinematics} have to be considered as lower limits
when compared to velocity dispersion measurements obtained for resolved 
objects (for which the seeing plays a much smaller role) at the $R_e$ (as done in Fig,~\ref{fig:mass-vdisp}). We note that the correction derived in \citet{Cappellari+06}, Eq.~1,  
has been defined and extensively tested only on resolved objects. 
In our case, given the extremely small sizes of the UCMGs, the \INSPIRE \, 2D spectra are 
fully seeing dominated. 
Hence, even if extracting from apertures encapsulating 50\% of the light, we do not have any spatial information on where this light comes from and hence we cannot apply that equation. 

\begin{figure}
    \centering
    \includegraphics[width=\columnwidth]{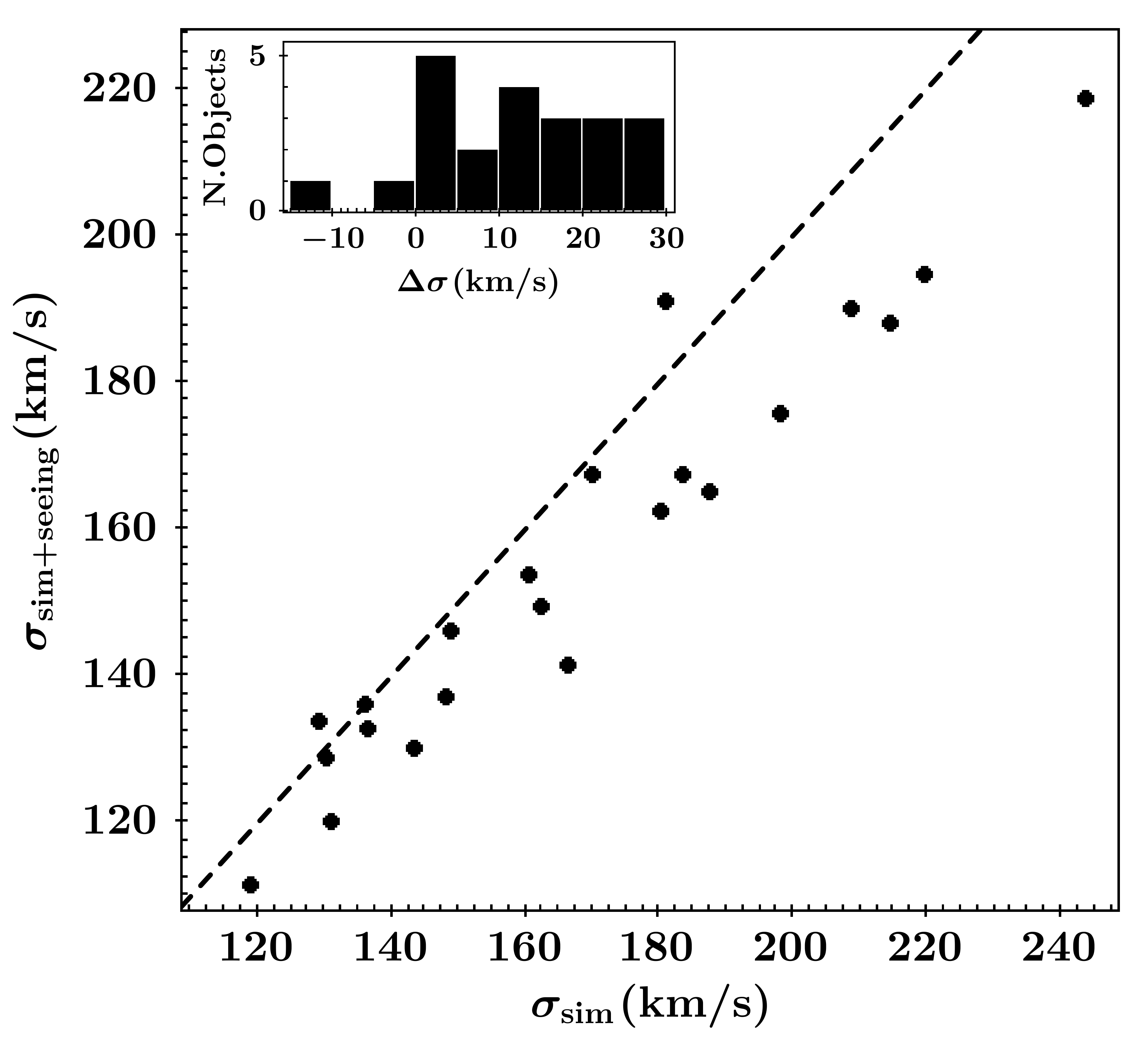}
    \caption{Velocity dispersion values measured for the simulated galaxies, without (x-axis) and with (y-axis) the seeing convolution. The histogram in the inset shows the difference between the $\sigma_{\text{sim}}$ and $\sigma_{\text{sim+seeing}}$, which is positive in 21/22 cases, and never larger than 30 km/s.}
    \label{fig:sigma_Re}
\end{figure}

In order to qualitatively estimate the variation on $\sigma_{\star}$ due to the seeing, we use simulated UCMGs from the IllustrisTNG cosmological simulations \citep{Springel18, Pillepich18, Naiman18, Marinacci18, Nelson18}. 
From the TNG100 run at $z=0.3$, we select galaxies with \Reff$<2$ kpc 
 and \Mstar$>6\times10^{10}$\Msun
 (the central value in the \INSPIRE\, DR1 redshift distribution) and retrieve 22 objects. We derive 2D projected sigma profiles by binning particles in apertures 
 of different sizes and calculate the dispersion with respect to the mean velocity in each bin, 
 weighting by the particle mass.  
 As expected, the velocity dispersion spatial profiles are decreasing in most of the cases. We then extract the $\sigma$ from an aperture containing 50\% of the light (i.e. the R50 for these simulated objects). Finally, we convolve these profiles with a Gaussian filter of width $1\arcsec$ ($4.6$ kpc at $z=0.3$) ---to simulate the effect of the seeing on the $\sigma$ profiles--- and recompute the velocity dispersion in the R50 aperture. 
The results of this test are shown in Figure~\ref{fig:sigma_Re}. On average, we find that $\sigma_{\text{seeing}} \sim \sigma \times 1.07$, i.e. the velocity dispersion inferred from a seeing-dominated spectra would be 1.07 times smaller than the real ones (only in 1/22 case $\sigma_{\text{sim+seeing}} > \sigma_{\text{sim}}$).

\section{Comparison with S20}
\label{app:s20compa}
A tentative estimate of the stellar velocity dispersion for 17 of the \INSPIRE\, DR1 objects was already presented in S20 where, however, it was inferred from spectra with very low S/N, hence the very large error bars. The velocity dispersion values reported in Tables 3 and 4 of \citetalias{Scognamiglio20} were extrapolated to the effective radii with Eq.~1 of \citet{Cappellari+06}.  Hence, although we do not believe that the correction is appropriate for ground-based spectra of UCMGs, where the seeing completely dominated, we apply the correction in the same way as in \citetalias{Scognamiglio20}, and bring our values to the \Reff too. 

\begin{figure}
\includegraphics[width=\columnwidth]{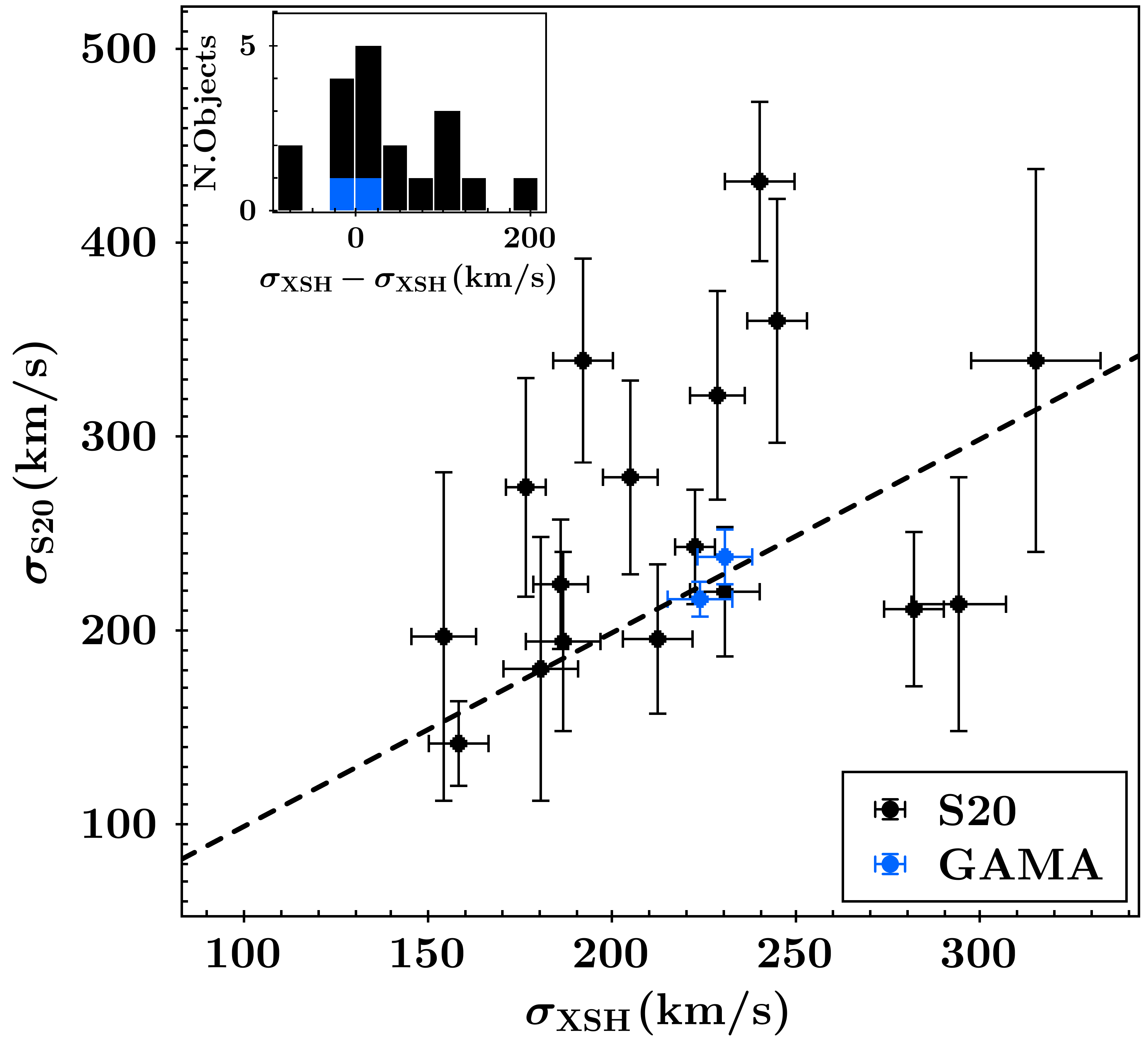}
\caption{Velocity dispersion values obtained from the R50 extracted spectra and then corrected to the \Reff\ compared to the one presented in S20 (in black) and obtained from GAMA spectra (in blue). All estimates are corrected to the \Reff using the same method. The histogram on the top shows the distribution of the difference between the two measurements.  }
\label{fig:comp_S20}
\end{figure}

Figure~\ref{fig:comp_S20} shows the values we obtain versus those measured in S20, both corrected in the same way to the same aperture.  
Here the agreement is suboptimal, although not terrible, as only five systems have stellar velocity dispersion measurements in disagreement for more than $1.5\sigma$ errors. 
Overall, the values reported in S20 are systematically larger than the ones we inferred, \text{as seen in the histogram shown in the inset of Figure~\ref{fig:comp_S20}}. Only for two galaxies do we find a slightly larger velocity dispersion value from the much better XSH spectra.

The remaining two objects (J0847+0112  and J0920+0212) have a publicly available GAMA spectrum. We downloaded them and  computed the velocity dispersion values using \ppxf\, with an identical configuration to that used for the XSH spectra. In this case (blue points), very good agreement is found between the values inferred from XSH and those inferred from GAMA spectra ($\sigma_{\mathrm{GAMA, J0847}} = 233\pm9$; $\sigma_{\mathrm{GAMA, J0920}} =239\pm14$). 

\section{Calculation of signal-to-noise ratio }
\label{app:signal_to_noise}

\begin{table*}
\centering
\begin{tabular}{lcccccccc|c}
\hline
\hline
  \multicolumn{1}{c}{ID} &
  \multicolumn{1}{c}{S/N$_{\text{OptEx}}$} &
  \multicolumn{1}{c}{S/N$_{R50}$} &
  \multicolumn{1}{c}{S/N$_{\text{OptEx}}$} &
  \multicolumn{1}{c}{S/N$_{R50}$} &
  \multicolumn{1}{c}{S/N$_{\text{OptEx}}$} &
  \multicolumn{1}{c}{S/N$_{R50}$} &
  \multicolumn{1}{c}{S/N$_{\text{OptEx}}$} &
  \multicolumn{1}{c}{S/N$_{R50}$}          &
  \multicolumn{1}{|c}{<S/N>}   \\
  \multicolumn{1}{c}{KiDS} &
  \multicolumn{1}{c}{UVB} &
  \multicolumn{1}{c}{UVB} &
  \multicolumn{1}{c}{VIS} &
  \multicolumn{1}{c}{VIS} &
  \multicolumn{1}{c}{CaHK} &
  \multicolumn{1}{c}{CaHK} &
  \multicolumn{1}{c}{MgFe} &
  \multicolumn{1}{c}{MgFe}  & 
  \multicolumn{1}{|c}{}\\
\hline
J0211-3155 & 10.6 & 8.4 & 26.8 & 18.0 & 40.8 & 27.0 & 40.8 & 27.9    & 25.0 \\
  J0224-3143 & 15.6 & 12.1 & 60.9 & 34.6 & 62.7 & 44.8 & 73.4 & 55.4 & 44.9 \\
  J0226-3158 & 17.8 & 13.6 & 40.3 & 24.9 & 50.0 & 35.0 & 29.5 & 29.7 & 30.1 \\
  J0240-3141 & 13.4 & 10.4 & 42.1 & 22.9 & 47.9 & 33.2 & 37.7 & 31.2 & 29.9 \\
  J0314-3215 & 15.5 & 12.3 & 51.3 & 28.7 & 47.2 & 34.5 & 32.8 & 34.2 & 32.1 \\
  J0316-2953 & 10.8 & 8.4 & 46.2 & 27.3 & 40.7 & 27.3 & 47.7 & 33.3  & 30.2 \\
  J0317-2957 & 15.5 & 10.7 & 38.3 & 19.9 & 45.2 & 31.0 & 36.4 & 29.4 & 28.3 \\
  J0321-3213 & 16.3 & 13.6 & 50.2 & 33.0 & 58.4 & 40.7 & 51.5 & 44.1 & 38.5 \\
  J0326-3303 & 15.7 & 11.9 & 46.9 & 26.9 & 47.8 & 33.0 & 43.4 & 36.3 & 32.7 \\
  J0838+0052 & 16.9 & 12.6 & 50.1 & 27.4 & 57.5 & 38.9 & 43.3 & 35.8 & 35.3 \\
  J0842+0059 & 9.3 & 8.2 & 28.4 & 17.4 & 35.8 & 27.4 & 26.3 & 25.8   & 22.3 \\
  J0847+0112 & 18.0 & 15.2 & 28.6 & 19.2 & 68.0 & 46.7 & 42.2 & 30.3 & 33.5 \\
  J0857-0108 & 12.3 & 9.6 & 29.8 & 19.8 & 37.6 & 24.7 & 30.3 & 24.6  & 23.6 \\
  J0918+0122 & 13.4 & 10.5 & 53.8 & 29.2 & 61.7 & 40.9 & 71.1 & 48.9 & 41.2 \\
  J0920+0212 & 12.7 & 9.1 & 35.6 & 18.9 & 50.3 & 32.1 & 46.3 & 30.8  & 29.5 \\
  J2305-3436 & 10.8 & 8.7 & 28.2 & 17.2 & 40.8 & 24.9 & 35.0 & 25.0  & 23.8 \\
  J2312-3438 & 23.6 & 18.3 & 99.8 & 60.9 & 64.2 & 43.5 & 77.2 & 55.4 & 55.4 \\
  J2327-3312 & 14.2 & 12.4 & 63.6 & 37.4 & 63.5 & 45.7 & 69.9 & 59.1 & 45.7 \\
  J2359-3320 & 12.0 & 9.9 & 34.3 & 21.5 & 43.3 & 22.8 & 35.4 & 21.8  & 25.1 \\
\hline\hline
\end{tabular}
\caption{S/N per Angstrom computed for the integrated UVB and VIS spectra and around some of the strongest stellar absorption features (the CaK and CaH doublet, $[3800-4000]$ \AA\, and the Mg and Fe lines $[5000-5500]$ \AA) of each galaxy for both extracting methods. \text{The last column list the mean value of the eight different estimates.} }
\label{tab:snr_calculation}
\end{table*}

\begin{figure}
\includegraphics[width=\columnwidth]{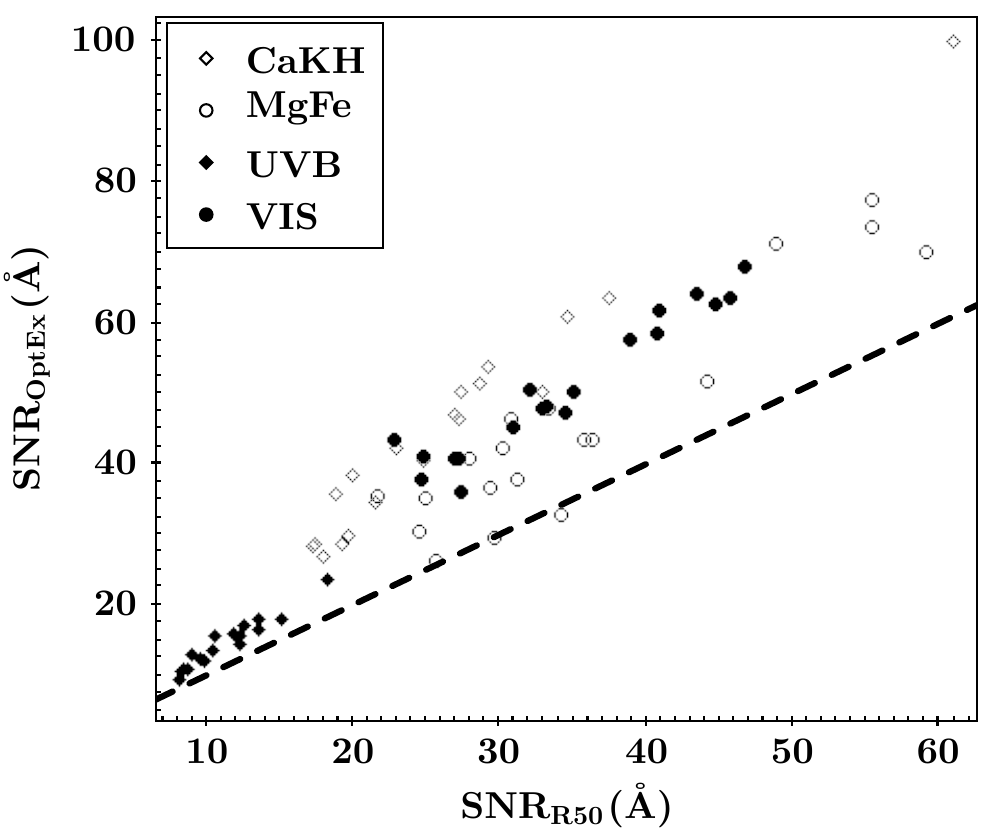}
\caption{Comparison between the S/N obtained from the R50 spectra and that from the OptEx ones in the two arms (integrated, filled symbols) and around the Ca H and K doublet in UVB (computed in the region $[3800-4000]$\AA, empty diamonds) and the MgFe lines in the VIS ($[5000-5500]$\AA, empty circles). The S/N obtained from the optimally extracted spectra is in all but two cases above the one-to-one relation, which is plotted as dashed black line. } 
\label{fig:snr_comparison}
\end{figure}

To calculate the S/N of the 1D spectra, we use the IDL code \texttt{}{DER\_SNR} \citep{Stoehr08} which estimates it directly from the flux, assuming that the noise is uncorrelated in wavelength bins spaced two pixels apart and that it is approximately Gaussian distributed.   
First, we bring all the spectra extracted with both methods, OptEx and R50, to their restframe wavelength; we then run the code considering four different regions: the entire UVB coverage, the entire VIS coverage (which slightly change from object to object according to their redshifts), the region from 3800 to 4000 \AA\, in the red part of the UVB arm ---where the strong calcium doublet is present (CaK3934.7, CaH3968.5)---, and the region from 5000 to 5500 \AA\, containing the Mgb5177 and several well-known iron stellar lines (Fe5270 and Fe5335). The resulting S/N values are all listed in Table~\ref{tab:snr_calculation}.

\end{document}